\documentclass[floatfix,reprint,amsmath,amssymb,superscriptaddress,aps,prx]{revtex4-1}
\usepackage[american]{babel}
\usepackage{graphicx}
\usepackage{bm}
\usepackage[hidelinks]{hyperref}
\usepackage[capitalize]{cleveref}
\usepackage[nice]{nicefrac}
\usepackage{physics}
\usepackage{xcolor}
\usepackage{xspace}
\usepackage{mathtools}  
\usepackage[normalem]{ulem} 
\usepackage{hyperref}
\usepackage{xcolor}
\hypersetup{colorlinks,bookmarksopen,bookmarksnumbered,
citecolor=teal,
linkcolor=teal,
pdfstartview=false,
urlcolor=teal}

\usepackage{orcidlink}

\begin{document}
\date{\today}

\title{Active Quantum Particles from Engineered Dissipation}

\author{Jeanne Gipouloux}
\affiliation{JEIP, UAR 3573 CNRS, Coll\`ege de France, PSL Research University, 11 Place Marcelin Berthelot, 75321 Paris Cedex 05, France}
\author{Matteo Brunelli}
\affiliation{JEIP, UAR 3573 CNRS, Coll\`ege de France, PSL Research University, 11 Place Marcelin Berthelot, 75321 Paris Cedex 05, France}
\author{Leticia F. Cugliandolo~\orcidlink{0000-0002-4986-8164}}
\affiliation{Sorbonne Universit\'e, Laboratoire de Physique Th\'eorique et Hautes Energies, CNRS-UMR 7589, 4 Place Jussieu, 75252 Paris Cedex 05, France}
\author{Rosario Fazio~\orcidlink{0000-0002-7793-179X}}
\affiliation{The Abdus Salam International Center for Theoretical Physics, Strada Costiera 11, 34151 Trieste, Italy}
\affiliation{Dipartimento di Fisica “E. Pancini”, Università di Napoli “Federico II”, Monte S. Angelo, I-80126 Napoli, Italy}

\author{Marco Schir\`o}
\affiliation{JEIP, UAR 3573 CNRS, Coll\`ege de France, PSL Research University, 11 Place Marcelin Berthelot, 75321 Paris Cedex 05, France}

\begin{abstract}
We introduce and characterize different models for an active quantum particle where activity arises from engineered dissipation-- specifically, from a suitably coupled nonequilibrium environment. These include a model of a particle moving on a lattice with coherent and dissipative hopping, as well as quantum generalizations of well-studied models of active behavior, such as the active Ornstein-Uhlenbeck process, run-and-tumble dynamics, and the active Brownian particle. Despite the different microscopic mechanisms at play, we show that all these models display key features of active motion. Notably, we observe a crossover from diffusive to active-diffusive behavior at long times, leading to an effective P\'eclet number, as well as a strong sensitivity to boundary conditions which, in our open quantum system context, arises  from the Liouville skin effect.  
We discuss the role of quantum fluctuations and experimental realizations with superconducting circuits or cold gases, closing with perspectives for many-body effects in quantum active matter.
\end{abstract}

\maketitle


Active matter describes a broad range of systems of relevance in physics, chemistry and biology~\cite{marchetti2013hydrodynamics,bechinger2016active,obyrne2022time,vrugt2025exactlyactivematter}. A key feature is their out-of-equilibrium nature and the fact that activity arises from the local injection of energy at the microscopic scale, which is partially converted into motion in the absence of external gradients.
Active particles display several unique features, from enhanced diffusivity to sensitivity to boundary conditions and disorder~\cite{granek2024colloquium}.
Interactions among active particles give rise to remarkable collective phenomena~\cite{vicsek1995novel,guillaume2004onset,tailleur2008statistical} and phase transitions~\cite{cates2015motility}, which can be described at coarse-grained scale by active field theories~\cite{wittkowski2014scalar,nardini2017entropy,tjhung2018cluster,granek2024colloquium}.
\begin{figure}[h!]
\centering
\includegraphics[width=0.98\linewidth]{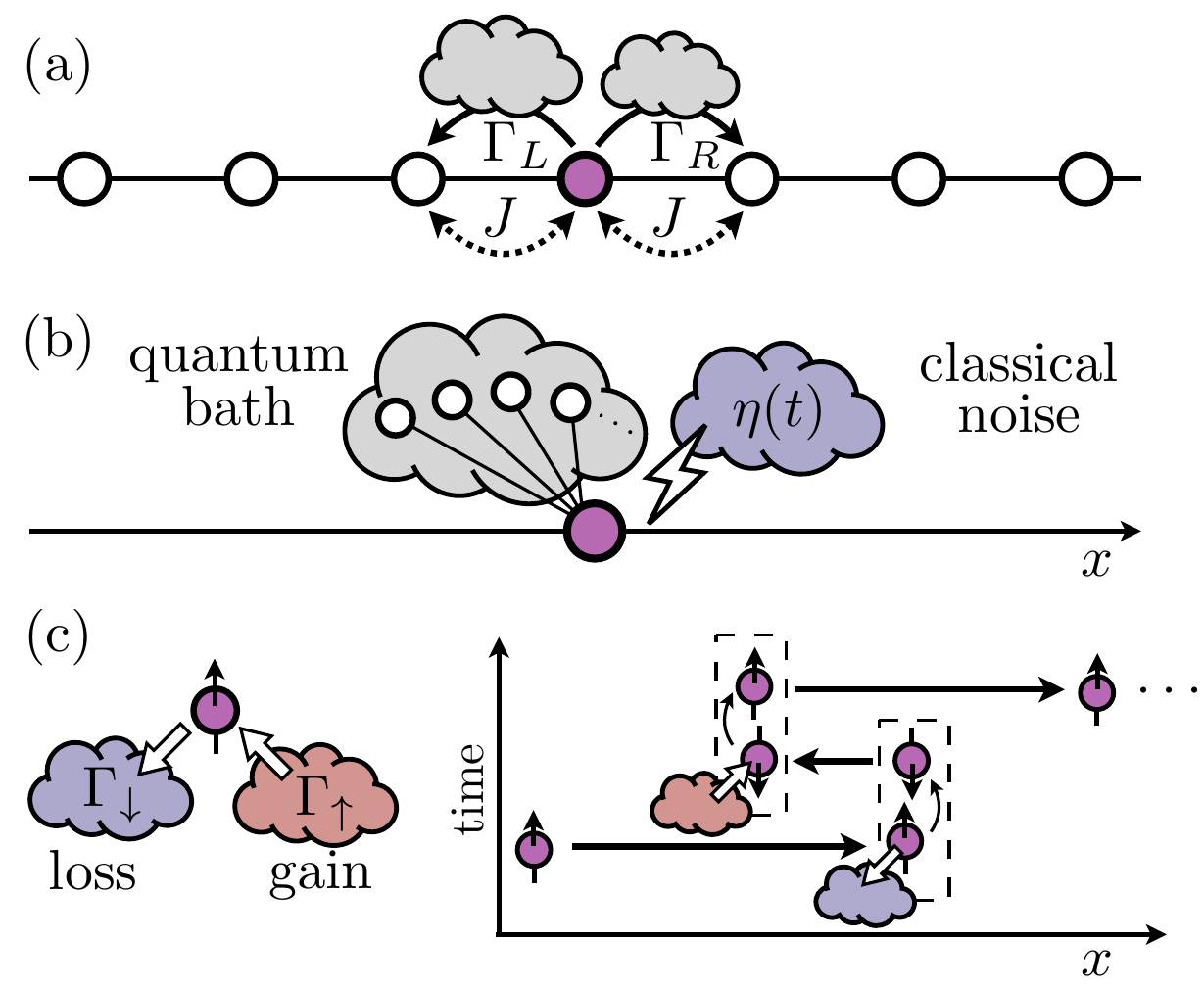}
\caption{Sketch of the setup. (a) A quantum particle moves in presence of coherent and environment-induced hoppings, with rates $J$ and $\Gamma_{L,R}$. (b) A dissipative quantum particle in a noisy force with finite persistence time. (c) A dissipative quantum particle coupled to a two-level system mimicking an effective spin-orbit coupling.}
\label{fig:sketch}
\end{figure}

An intriguing question is whether active behavior is unique to classical systems or can manifest in the quantum world as well. 
Despite the growing interest in out-of-equilibrium quantum systems over the past few decades--encompassing both unitary~\cite{polkovnikov2011colloquium} and driven-dissipative  
dynamics~\cite{landi2022nonequilibrium,fazio2025many,sieberer2025universality}--only recently have there been first attempts to address this question. 
Different models of a quantum active particle have been proposed, based on non-unitary quantum walks~\cite{yamagishi2024proposal}, an external trapping potential mimicking active behavior~\cite{antonov2025engineering,antonov2026modelingdissipationquantumactive}, and spin–orbit coupling combined with a mechanism for heat-to-motion conversion~\cite{penner2025heattomotionconversionquantumactive}. The dynamics of a monitored qubit has been shown to display features of active behavior~\cite{kundu2025runtumbledynamicsbiased}. Collective effects in interacting quantum active systems have also started to be discussed~\cite{khasseh2024activequantumflocks,Adachi_2022_activity-inducedquantum, Takasan_2024_activity-inducedquantum,nadolny2024nonreciprocalsynchronizationactivequantum,vonOppen}.

Several fundamental questions persist regarding the manifestation of active behavior in quantum systems and its connection to the classical phenomenology.
\begin{figure*}[t]
    \centering
    \includegraphics[width=1\linewidth]{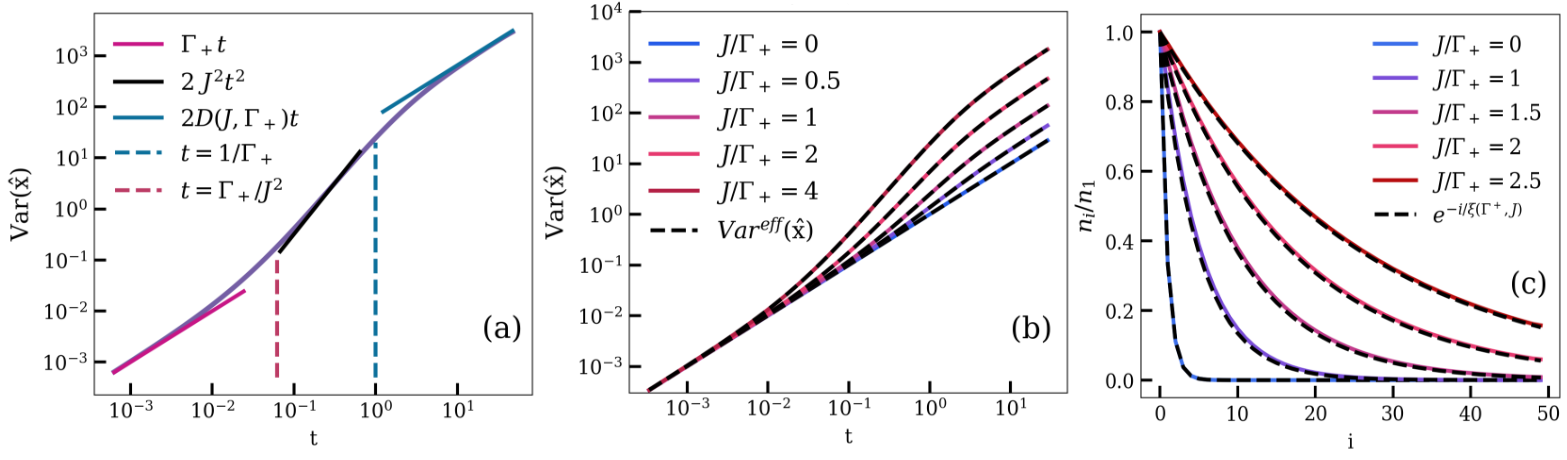}
\caption{Environment-assisted hopping model. (a) Variance of the particle position for $J/\Gamma_+ = 4$ and $\Gamma_-=0$ displaying a crossover from diffusive to ballistic to active-diffusive behavior, the latter 
regime with enhanced diffusion coefficient $D(J,\Gamma_+)$. The two crossover times are shown with vertical dashed lines. (b) Crossover in the variance for different values of $J/\Gamma_+$.  The black dashed lines represent the formula~(\ref{eq:varX-exact-modela}). (c) Steady-state density profile for open-boundary conditions, for different values of $J/\Gamma_+$ with $\Gamma_- = 0.5$ - the dotted lines represent  the analytic expression~\eqref{eq:SkinCorr}. In the three panels $\Gamma_+ = 1$. 
}
\label{fig1}
\end{figure*}
The self-propulsion of classical active particles--modeled either as a fluctuating velocity or a fluctuating active force--accelerates their dynamics relative to the passive case and enhances long-time diffusivity. This behavior is found in all models of active matter, such as the run-and-tumble dynamics~\cite{berg1972chemotaxis,LOVELY1975477,schnitzer1993theory}, active Brownian motion~\cite{fily2012athermal,Romanczuk2012active}, and the active Ornstein-Uhlenbeck process~\cite{hanggi1994colored,szamel2014self,martin2021statistical}. 
In this Letter, we introduce and characterize minimal models of an {\it active quantum particle} that (i) are directly inspired by those studied in the classical literature and share their basic phenomenology, and (ii) display emergent activity due to the interplay between coherent quantum dynamics and engineered dissipation. This involves coupling to a suitably designed environment, which allows energy to be dissipated while at the same time being converted into motion. First, we consider a particle moving via coherent and environment-induced hopping [Fig.~\ref{fig:sketch}(a)], which  displays active dynamics due to virtual dissipative processes induced by the coherent hopping.
We then introduce a model of a dissipative quantum particle subject to a noisy force with finite persistence time--a quantum active Ornstein-Uhlenbeck process (qAOUP)
[Fig.~\ref{fig:sketch}(b)]. In this case, activity is introduced classically and competes with quantum Brownian dynamics at low temperatures. Finally, we introduce a quantum run-and-tumble dynamics (qRTD) or quantum active Brownian particle (qABP) [Fig.~\ref{fig:sketch}(c)], where activity arises from the coupling between motion and a dissipatively-controlled internal two-level system, mimicking velocity fluctuations.
Despite the different microscopic mechanisms 
at play, all the models share fundamental aspects of active behavior. Notably, the variance of the particle position displays a dynamical crossover from a short-time passive regime to a long-time active regime, characterized by enhanced diffusivity mediated by an intermediate ballistic scaling. We highlight the role of quantum fluctuations and potential experimental implementations using ultracold atoms or superconducting circuits.
\emph{Environment-assisted hopping model---}We first consider a single particle hopping 
on a chain with Hamiltonian $\hat{\mathcal{H}}=J\sum_{i=1} [\hat{c}_{i}^{\dagger}\hat{c}_{i+1}+\hat{c}_{i+1}^{\dagger}\hat{c}_{i} ]$, where $J$ is the coherent hopping rate. In addition to unitary evolution, the particle evolves dissipatively via incoherent hopping processes, see Fig.~\ref{fig:sketch}(a). The density operator, $\hat \rho$, evolves according to the Lindblad master equation~\cite{fazio2025many}
\begin{equation}\label{eqn:lindblad}
\partial_{t} \hat\rho=-i \, [\mathcal{\hat{H}}, \hat\rho ]+\!\!\! \sum_{i,\mu=L,R}
\! \Big( 
\hat{L}_{i,\mu} \hat\rho \hat{L}_{i,\mu}^{\dagger}
\! -  
\dfrac{1}{2} \, \{ \hat{L}_{i,\mu}^{\dagger}\hat{L}_{i,\mu},\hat\rho \} 
\Big)\,,
\end{equation}
where  $\hat L_{i,\mu}$ are a set of jump operators describing incoherent hopping with rates $\Gamma_{L}$ and $\Gamma_R$, i.e.
$\hat L_{i,L} = \sqrt{\Gamma_L}\, \hat c^\dagger_i \hat c_{i + 1}$ and $\hat L_{i,R} = \sqrt{\Gamma_R} \, \hat c_{i + 1}^\dagger \hat c_i$. These jumps induce dephasing, but crucially act across a bond rather than on-site. 
They have been considered to model dissipative transport~\cite{garbe2024bosonic, Minoguchi_2025_dissipative} and in connection with the quantum stochastic exclusion process~\cite{Temme_2012,bernard2019open,hruza2023coherent}. Combined with coherent hopping, as in Eq.~\eqref{eqn:lindblad}, they have been discussed in relation to Liouvillian skin effect~\cite{haga2021liouvillian}, in spinless fermions~\cite{penc2025linearresponseexacthydrodynamic} and  in a  bosonic dimer~\cite{Solanki_2025_dimer}.
In the absence of dissipation, Eq.~\eqref{eqn:lindblad}  describes the ballistic motion of a free quantum particle, while 
for $J=0$ one recovers the diffusive behavior of a classical passive particle. 
Remarkably, the interplay between the two types of hopping processes results in an effective active behavior.
We note that the Lindblad dynamics with the local jump operators above do not satisfy detailed balance, preventing thermalization to Gibbs equilibrium, and that the total particle number is conserved. 

In the following, we focus on the single particle sector of Eq.~(\ref{eqn:lindblad}) and we numerically compute  the variance of the particle position, 
$\mbox{Var}(\hat{x}) = \langle \hat{x}^2\rangle - \langle \hat{x}\rangle^2$ with $\langle\dots\rangle=\mbox{Tr}(\hat \rho\dots)$,
starting from a wave packet localized in the middle of the chain.
For periodic boundary conditions,  $\mbox{Var}(\hat{x})$ is only controlled by the average dissipation $\Gamma_+=\Gamma_L+\Gamma_R$~\cite{supplementary_mat}. We first consider the symmetric case, $\Gamma_L=\Gamma_R=\Gamma_+/2$.
In Fig.~\ref{fig1}(a) we plot the variance, which displays a crossover from diffusive behavior at short times, $\mbox{Var}(\hat{x}) = \Gamma_+ \, t$, to an increased diffusion at longer times, $\mbox{Var}(\hat{x})= 2D(J,\Gamma_+) \, t$, via a transient ballistic regime where $\mbox{Var}(\hat{x})=2J^2t^2$. Numerically, we estimate the parameter dependence of the diffusion coefficient in the last regime as \begin{align}\label{eq:D}
D(J,\Gamma_+)=\frac{\Gamma_+}{2}\left(1+ \frac{4J^2}{\Gamma_+^2}\right)\,.
\end{align}
As we show in Fig.~\ref{fig1}(a)-(b), the first crossover occurs at
$t^*_1\sim\Gamma_+/J^2$ while the second time scale is $t^*_2= 1/\Gamma_+$
(we assume $\Gamma_+^2 < 2J^2 $ so that $t^*_1< t^*_2$). To gain further understanding of this result we write down the Lindblad master equation in the single particle basis and assume that, at least for moderate $J/\Gamma_+$, the density matrix is almost diagonal, i.e. quantum coherences far away from the diagonal decay fast to zero. Using this approximation, we obtain a closed expression for the particle variance~\cite{supplementary_mat}
\begin{equation}
\mbox{Var}^{\raisebox{0.09cm}{\!\!\!\!\!\!\! \scriptsize \rm eff}}(\hat{x})=\Gamma_+ t+\frac{4J^2}{\Gamma_+}t+\frac{4J^2}{\Gamma_+^2}\left(e^{-\Gamma_+ t}-1\right)
\; . 
\label{eq:varX-exact-modela}
\end{equation}
This analytical result is compared with the numerics in Fig.~\ref{fig1}(b). 
Surprisingly enough, this function captures the first diffusive regime
at $t\ll \Gamma_+/J^2 \ll 1/\Gamma_+$,  
the ballistic one 
at $\Gamma_+/J^2 \ll t \ll 1/\Gamma_+$, 
the final diffusive one at $ 1/\Gamma_+\ll t$, 
and matches the data throughout the two crossovers. More importantly, the functional  form is strongly reminiscent of the one of active classical particles~\cite{supplementary_mat}: it displays the expected crossover from diffusive to ballistic to activity-enhanced diffusive motion, with an enhancement of the diffusion coefficient that we can parametrize via an effective P\'eclet number~\cite{bechinger2016active}, $D(J,\Gamma_+) \equiv D\left(1+\mbox{Pe}^2\right)$, with  $D=\Gamma_+/2$ and $\mbox{Pe}=2J/\Gamma_+$,  a signature of active behavior~\cite{bechinger2016active,cugliandolo2025introductionlangevinstochasticprocesses,supplementary_mat}. In the present context, the crossover can be understood as the interplay of two diffusive processes, a fast one due to dissipation $\Gamma_+$ and a second one due to coherent hopping-induced dissipation, setting in for times $t\gg \Gamma_+$, see Eq.~(\ref{eq:varX-exact-modela}). 

In the presence of asymmetric dissipation, i.e., when $\Gamma_-=\Gamma_L-\Gamma_R >0$ further interesting aspects of our model, namely its sensitivity to boundary conditions, come into play. Specifically, we discuss the dynamics of the quantum particle on a lattice with open boundaries, such that when reaching the edges the particle can only hop back into the chain~\cite{supplementary_mat}.
When initialized at the center of the chain, the particle propagates towards the boundary with a non-zero velocity, $\langle \hat{x}\rangle\sim \Gamma_- t$ and diffusive fluctuations, and localizes around it.
In Fig.~\ref{fig1}(c) we plot the steady-state profile of the particle density $\langle n_i\rangle$, displaying an exponential localization at the edge of the system~ $\langle n_{i}\rangle/\langle n_1\rangle = \xi^{-1} \, e^{-i/\xi}$ with a localization length $\xi\ll L$ that increases with $J/\Gamma_+$.
This localization is a manifestation of the Liouville skin-effect~\cite{haga2021liouvillian}. Indeed, for $J=0$ the Lindblad operator can be mapped to an effective Hatano-Nelson Hamiltonian, which shows a skin effect with a localization length $\xi=-1/\log(\Gamma_R/\Gamma_L)$~\cite{haga2021liouvillian}. To understand our result at finite $J/\Gamma_+$ we 
project away the off-diagonal elements of the density matrix~\cite{supplementary_mat}. This allows us to derive 
\begin{equation}
    \xi= -1/\log(\frac{ D(J,\Gamma_+)-\Gamma_-/2}{D(J, \Gamma_+) + \Gamma_-/2 } ) 
    \simeq \frac{D(J,\Gamma_+)}{\Gamma_-}
    \; , 
    \label{eq:SkinCorr}
\end{equation}
with $D(J,\Gamma_+)$ the enhanced diffusion coefficient of Eq.~(\ref{eq:D})
(we took $\Gamma_->0$ for definiteness). Remarkably, this expression perfectly matches our numerics, see Fig.~\ref{fig1}(c), even for values of $J/\Gamma_+\sim 1$. This behavior is reminiscent of the dynamics of an active particle which, when confined in a box, sticks to the boundary and localizes on a scale set by the ratio between diffusion and drift~\cite{palacci2010sedimentation,razin2020entropy}. 
\begin{figure*}[t]
    \centering
    \includegraphics[width=\linewidth]{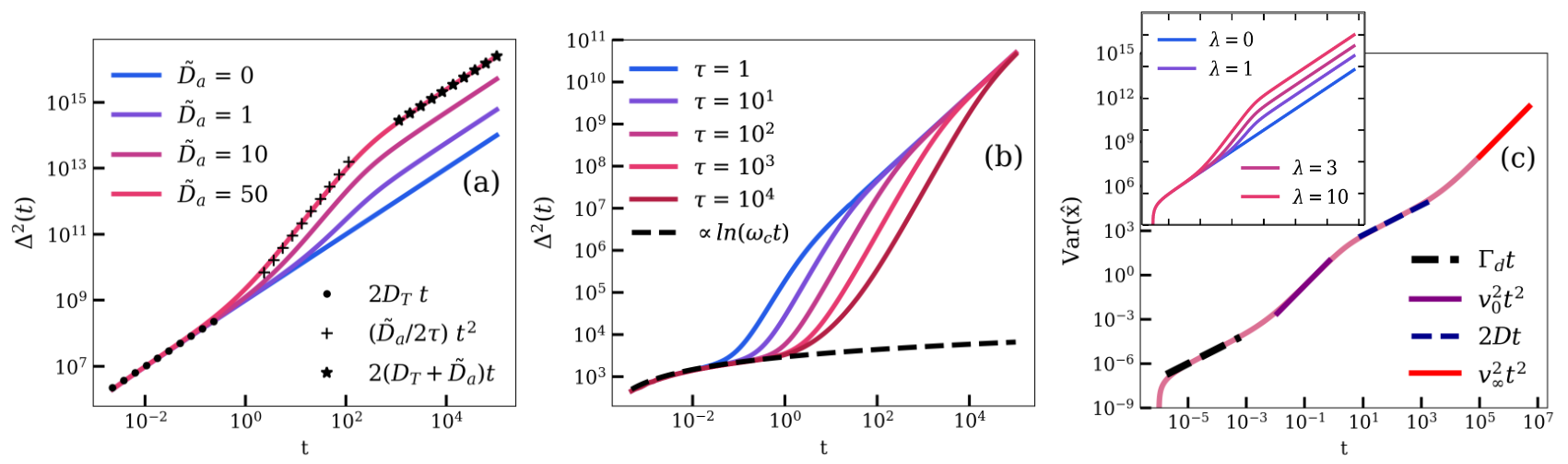}
\caption{(a)-(b) Dynamics of the mean-squared displacement $\Delta^2(t)$ for the qAOUP. (a) Diffusive-ballistic-diffusive crossover at finite temperature, obtained by changing the active diffusion coefficient $\tilde{D}_a=D_a/4\gamma^2$ at fixed persistence time $\tau$. Numerical Parameters: $\gamma=10^{-5}$, 
$\beta=1/T=10^{-4}$, $\tau=100$, $D_T=T/2\gamma$. (b) Zero temperature crossover from quantum Brownian motion to active diffusion, tuning the persistence time $\tau$. Numerical Parameters: $\gamma=10^{-3}$, $\beta=1$, $D_a=1$. In both (a)-(b) panels the frequency cut-off is $\omega_c = 10^4$. (c) Dynamics of the particle variance $\mbox{Var}{(\hat{x})}$ for the qABP model for symmetric rates, showing the crossover from diffusive to active diffusive motion, with $D$ given in Eq.~(\ref{eqn:D_qabp}). The velocities $v_0,v_{\infty}$ of the intermediate and long-time ballistic motion are given in Ref.~\cite{supplementary_mat}. Numerical Parameters: $m=10$, $ \lambda, \Gamma_+, \Gamma_d$ = 5,\, 1, \, 0.1.
Inset: infinite mass limit for different values of the spin-orbit coupling $\lambda$.}
\label{fig:qABM,qAOUP}
\end{figure*}

Enhanced diffusion and boundary localization demonstrate how the interplay between coherent dynamics and dissipation-induced hopping 
produces characteristic features of active particle behavior. We next extend this analogy by introducing two quantum models inspired by well-established classical active particle dynamics.

\emph{ Quantum Active Ornstein-Uhlenbeck Particle---}In a classical AOUP, activity arises from the coupling to a noisy environment, 
such as an active force with a finite persistence time, which breaks the fluctuation-dissipation theorem (FDT). We consider a quantum particle with coordinate and momentum operators $[\hat{x},\hat{p}]={\rm i}$, coupled to a quantum bath of harmonic oscillators $\hat{x}_\alpha$ and exposed to classical colored (non-Markovian) Gaussian noise $\eta(t)$. The Hamiltonian of the system reads 
\begin{align}\label{eqn:HqAOUP}
\hat{\mathcal H}_{\rm qAOUP} =\frac{\hat{p}^2}{2m} +\hat{\mathcal H}_{\rm bath}(\hat{x})+g\hat{x} \eta(t)\,,
\end{align}
where $\hat{\mathcal H}_{\rm bath}$ describes coupling to regular Ohmic dissipation~\cite{caldeira1981influence,Weiss:2021uhm,kamenev2023field}
\begin{align}
\hat{\mathcal H}_{\rm bath}(\hat{x})=\sum_{\alpha}\left(\frac{\hat p_{\alpha}^2}{2m_{\alpha}}+\frac{m_{\alpha}\omega_{\alpha}^2}{2}\hat x_{\alpha}^2\right)+\hat{x}\sum_{\alpha} g_{\alpha}\hat{x}_{\alpha}\,,
\end{align}
with spectrum $J(\omega)=\sum_{\alpha} (g_{\alpha}^2/m_{\alpha}\omega_{\alpha})\, \delta(\omega-\omega_{\alpha})= 4\gamma \omega\exp(-\omega/\omega_c)$ and $\gamma$ the associated friction. The noise $\eta(t)$ is characterized by a finite persistence time, i.e. 
$\overline{\eta(t)\eta(t')}= (D_a/\tau) \, \exp(-\vert t-t'\vert/\tau)$ as for a classical AOUP~\cite{martin2021statistical,supplementary_mat}. By writing the Keldysh 
action~\cite{kamenev2023field}, 
the quantum dynamics of 
$\hat x$ are shown to map onto the ones of an AOUP in the semiclassical $\hbar\to 0$ 
limit~\cite{supplementary_mat}.
For finite $\hbar$ the dynamics of QAOUP are controlled by the interplay between activity, 
thermal and quantum fluctuations. We focus on the stationary regime 
and compute the mean-squared displacement $\Delta^2(t)=\overline{\langle \left(\hat{x}(t)-\hat{x}(0)\right)^2\rangle}$ that we write in the overdamped limit $m/\gamma\ll 1$ as~\cite{supplementary_mat}
\begin{align}
\Delta^2(t)= \int \frac{d\omega}{2\pi}
\frac{(1-\cos (\omega t))\left(\mbox{Im}\Sigma^K(\omega)+4\mathcal{D}(\omega)\right)}{4\gamma^2\omega^2} \,,
\end{align}
where $\mbox{Im}\Sigma^K(\omega)=J(\omega) \coth (\beta\omega/2)$ is the Keldysh self-energy (noise) of the quantum bath at temperature $T=1/\beta$, while ${\mathcal D}(\omega)$ is the kernel of the classical noise, ${\mathcal D}(\omega)=D_a/(1+\tau^2\omega^2)$. We first consider a finite temperature bath, leading to a Keldysh component which saturates to a finite value at zero frequency, $\Sigma^K(\omega\rightarrow0)=8\gamma T$. At short times, $t\ll \tau$ or $\omega\tau\gg 1,$ the active noise is irrelevant and the dynamics are dominated by standard thermal diffusion $\Delta^2(t)=2 D_T \, t$, with $D_T=T/2\gamma$. At long-times, the classical noise sets in and enhances diffusion, $\Delta^2(t)= 2(D_T+\tilde{D}_a) \, t$, with $\tilde{D}_a=D_a/4\gamma^2$.  The crossover between the two diffusive regimes is ballistic, as shown in Fig.~\ref{fig:qABM,qAOUP}(a), and sets in at $t^{*}\sim (D_T/\tilde{D}_a)\tau$, as in classical AOUPs~\cite{supplementary_mat}. For a zero temperature bath, instead, the quantum noise spectrum becomes $\Sigma^K(\omega)\sim {\rm i} \gamma\vert\omega\vert$, leading to long-ranged power-law noise-noise correlations in time. As a result, $\Delta^2(t)$ grows logarithmically at short times, a signature of quantum Brownian motion~\cite{hakim1985quantum,fisher1985quantum,Weiss:2021uhm}, while at longer times the active bath becomes relevant and leads to active diffusion. The crossover scale to active behavior is $t^{\star}\sim \sqrt{(\tau/\tilde{D}_a)\mbox{log}(\omega_c\tau/\tilde{D}_a)}$ which is parametrically smaller than in the classical AOUP case, i.e. a dissipative quantum particle reacts faster to an active stimulus than its classical counterpart. Using the Green's functions formalism and FDT~\cite{Cugliandolo_2011}, we derive the effective temperature $T_{\rm eff}(\omega)=T+D_a/[(2\gamma) \, (1+\tau^2\omega^2)]$.  At low frequencies, the effective temperature is enhanced by activity, in agreement with the result of the diffusion coefficient. We note that even in the limit $\tau\rightarrow 0$, corresponding to a fast active bath, $T_{\rm eff}(\omega)$ remains finite as opposed to a quantum particle evolving under Lindblad dynamics which heats up to infinite temperature due to the absence of friction~\cite{supplementary_mat}.

\emph{Quantum Run-and-Tumble Dynamics and Quantum Active Brownian Motion---}In the classical Run-and-Tumble dynamics (RTD) and Active Brownian Particle (ABP), 
activity emerges because the particle velocity fluctuates in time~\cite{supplementary_mat}, as telegraphic or Gaussian noise, respectively. To mimic these fluctuations we consider a quantum particle coupled to an internal degree of freedom, modeled as a two-level system (TLS), which is dissipatively manipulated. 
We take the Hamiltonian to be 
\begin{align}
\hat{\mathcal H}=\frac{\hat{p}^2}{2m}+\lambda \, \hat{p}\hat{\sigma}_z \,,
\end{align}
where $\lambda$ plays the role of an effective spin-orbit coupling~\cite{dalibard2011colloquium}. Dissipation is provided by the jump operators $\hat L_{\uparrow}=\sqrt{\Gamma_{\uparrow}} \, \hat \sigma_{+}$ and $\hat L_{\downarrow}=\sqrt{\Gamma_{\downarrow}} \, \hat \sigma_{-} $ describing pump/losses on the TLS and by $\hat L_d=\sqrt{\Gamma_d} \, \hat p$, leading to position diffusion for the particle.
Unlike Ref.~\cite{penner2025heattomotionconversionquantumactive}, we do not impose any temperature bias difference on the TLS, our particle being out of equilibrium because translational and internal degrees of freedom are coupled to different baths. The dynamics of this model can be solved in closed form~\cite{supplementary_mat}. The particle's momentum is conserved by the dynamics, and we set it to $\langle \hat p(t)\rangle=\langle \hat p(0)\rangle$. The variance of the particle position satisfies the equation
\begin{align}\label{eqn:dvarXdt}
\frac{d\mbox{Var}(\hat x)}{dt}=\frac{2}{m}\mbox{Cov}(\hat x,\hat p)+2\lambda\mbox{Cov}(\hat \sigma_z,\hat x)+\Gamma_d
\end{align}
where $\mbox{Cov}(\hat A,\hat B)=\langle \hat A\hat B\rangle-\langle \hat A\rangle \langle \hat B\rangle$. The first term in Eq.~(\ref{eqn:dvarXdt}) describes ballistic motion due to inertia, the second one introduces activity while the last one leads to diffusion. For $\lambda=0$ the particle performs a passive motion with a crossover from diffusive dynamics at short-times $\mbox{Var}(\hat x)=\Gamma_d t$ to ballistic motion on times $t_{\rm m}\sim \Gamma_d m^2/\mbox{Var}(\hat p)_0$, that we take to be the largest scale in the problem. For $\lambda\neq0$ we consider first the case of symmetric rates $\Gamma_{\uparrow}=\Gamma_{\downarrow}=\Gamma_+/2$, starting from a factorized initial condition, $\mbox{Cov}(\hat \sigma_z,\hat x)_0=0$. The dissipative dynamics leads to a build-up of correlations between particle and TLS, with $\mbox{Cov}(\hat \sigma_z,\hat x)\rightarrow \lambda/\Gamma_+$ on time scales $t\gg 1/\Gamma_+$, leading to an enhancement of diffusion $\mbox{Var}(\hat{x})=2Dt$ for $1/\Gamma_+\ll t \ll t_{\rm m} $, with 
\begin{align}\label{eqn:D_qabp}
D=\frac{\Gamma_d}{2}\left(1+\frac{2\lambda^2}{\Gamma_d\Gamma_+}\right)\,.
\end{align}
The crossover between the two diffusive scaling regimes is ballistic and controlled by the effective spin-orbit coupling $\lambda$, as shown in Fig.~\ref{fig:qABM,qAOUP}(c), which also shows the onset of ballistic motion at longer times due to inertia. The associated P\'eclet number from Eq.~(\ref{eqn:D_qabp}) reads $\mbox{Pe}=\sqrt{2}\lambda/\sqrt{\Gamma_d\Gamma_+}$, a form reminiscent of RTD and ABP~\cite{supplementary_mat}.
For asymmetric rates $\Gamma_-=\Gamma_{\uparrow}-\Gamma_{\downarrow}\neq 0$ the qualitative behavior is similar, with the asymmetry reducing the active diffusion enhancement, as in RTD~\cite{supplementary_mat}. Furthermore, the particle's velocity acquires a contribution proportional to the steady-state population of the TLS 
$\langle \hat \sigma_z\rangle_{\infty}=\Gamma_-/\Gamma_+$ achieved for $t\gg1/\Gamma_+$, i.e.  $\langle \hat x\rangle\sim vt$, with $v=\langle \hat p(0)\rangle/m+\lambda\Gamma_-/\Gamma_+$, which vanishes if the system is in thermal equilibrium~\cite{supplementary_mat}. 

To explicitly link our model to classical active stochastic dynamics, we employ a quantum trajectory unraveling of the Lindblad master equation, which describes the system’s conditional evolution under continuous monitoring~\cite{wiseman2009quantummeasurementand,landi2024current} (see also~\cite{viotti2026unpublished}).
In particular, by performing homodyne detection of all dissipative channels,
corresponding to Brownian noises, the resulting stochastic dynamics takes a form similar to the ABP. On the other hand, if the TLS is monitored with photo-counting, quantum jump events become Poisson distributed and the TLS evolves according to a telegraph noise, leading to RTD~\cite{supplementary_mat}. Finally, we note that a dissipative version of spin-motion coupling gives rise to similar active behavior for particle momentum rather than position~\cite{supplementary_mat}.

\emph{Discussion---}The models we discussed can be readily implemented experimentally in a variety of platforms.
The combination of coherent and dissipative hopping needed in the first model can naturally arise in ultracold atoms in optical lattices, where laser assisted hopping gives rise to the necessary incoherent processes~\cite{laflamme2017continuous,pichler2010nonequilibrium,haga2021liouvillian}. 
The implementation of qAOUP is natural using superconducting circuits and Josephson junctions, where the dynamics of the phase across the junction in a dissipative RC circuit performs a quantum Brownian motion~\cite{H.Grabert_1998,J.Ankerhold_2004,berthold2017quantum,cattaneo2021engineering,Weiss:2021uhm}, which can be supplemented by a colored non-Markovian noise describing a fluctuating voltage~\cite{dallatorre2010quantum,dallatorre2012}. Furthermore, physics similar to the qAOUP could also be engineered with other quantum simulation platforms, such as impurities in cold atoms~\cite{bonart2012from,petkovic2023dissipative} or optomechanics~\cite{carmele2020quantum}, in the presence of additional colored noise. Finally, for qRTD and qABP, the necessary spin-to-motion coupling can be engineered in atomic physics, where it describes a simple two-level atom with motional degrees of freedom in the presence of a synthetic spin-orbit coupling~\cite{lin2011spin,dalibard2011colloquium,galitski2013spin,chalopin2020probing}.

\emph{Conclusions -- } In this work we introduced and studied three models of an active quantum particle, all displaying a crossover from passive to active dynamics with enhanced diffusion due to fluctuations. The microscopic mechanisms responsible for the emergence of active behavior are different in the three cases, even though the common theme is the role of engineered dissipation as a resource for energy-to-motion conversion. In the environment-assisted hopping model, we have shown that hopping-induced dissipation, via quantum fluctuations of coherent hopping, is responsible for the dynamical crossover and the enhancement of diffusion at long times. In qAOUP on the contrary, activity is introduced classically via a noise with finite correlation time, which competes with the logarithmically slow propagation of quantum Brownian motion and gives rise to active diffusion at long-times. Finally, in qRTD and qABP, active behavior is introduced via coupling to an internal degree of freedom, which is dissipatively manipulated and emulates velocity fluctuations. Here, the dynamical crossover and the enhancement of diffusion set in after the motional and internal degrees of freedom become correlated. 

The models introduced here are inspired by standard models of classical active particles and offer ideal platforms to explore similarities and differences due to quantum mechanics. Promising directions for future work include the study of how quantum active particles respond to external potentials and boundaries, leading to competition between localization and active fluctuations. Moreover, it is natural to consider  many-body models built out of our quantum active particles and search for motility-induced phase transitions, flocking,  and other collective phenomena of the resulting quantum active matter.

\emph{Acknowledgements---} We thank L. Arrachea, S. Nascimbene, L. Viotti, F. von Oppen  for inspiring discussions. M.B. and M.S. acknowledge funding from the European Research Council (ERC) under the European Union's Horizon 2020 research and innovation program (Grant agreement No. 101002955 -- CONQUER). We acknowledge Coll\`{e}ge de France IPH cluster where the numerical calculations were performed.  L.F.C. acknowledges funding from ANR-20-CE30-0031 THEMA. R.F. acknowledges funding from the European Research Council (ERC), Grant agreement No. 101053159 -- RAVE.


%

\end{document}


\date{\today}
	
	\title{Supplemental Material to `Active Quantum Particles from Engineered Dissipation'}
	
\author{Jeanne Gipouloux}
\affiliation{JEIP, UAR 3573 CNRS, Coll\`ege de France, PSL Research University, 11 Place Marcelin Berthelot, 75321 Paris Cedex 05, France}
\author{Matteo Brunelli}
\affiliation{JEIP, UAR 3573 CNRS, Coll\`ege de France, PSL Research University, 11 Place Marcelin Berthelot, 75321 Paris Cedex 05, France}
\author{Leticia F. Cugliandolo~\orcidlink{0000-0002-4986-8164}}
\affiliation{Sorbonne Universit\'e, CNRS-UMR 7589 Laboratoire de Physique Th\'eorique et Hautes Energies, 4, Place Jussieu, 75252 Paris Cedex 05, France}
\author{Rosario Fazio~\orcidlink{0000-0002-7793-179X}}
\affiliation{The Abdus Salam International Center for Theoretical Physics, Strada Costiera 11, 34151 Trieste, Italy}
\affiliation{Dipartimento di Fisica “E. Pancini”, Università di Napoli “Federico II”, Monte S. Angelo, I-80126 Napoli, Italy}

\author{Marco Schir\`o}
\affiliation{JEIP, UAR 3573 CNRS, Coll\`ege de France, PSL Research University, 11 Place Marcelin Berthelot, 75321 Paris Cedex 05, France}

\maketitle

\onecolumngrid

\renewcommand{\thefigure}{S\arabic{figure}}
\renewcommand*{\citenumfont}[1]{S#1}
\renewcommand*{\bibnumfmt}[1]{[S#1]}

\newcounter{ssection}
\stepcounter{ssection}

\setcounter{table}{0}
\setcounter{page}{1}
\setcounter{figure}{0}
\setcounter{equation}{0}

\makeatletter
\renewcommand{\theequation}{S\arabic{equation}}
\tableofcontents


\section{Models of Classical Active Particles}\label{sec:classical}

In this Section we recall a few properties of standard models of classical active particles. The run-and-tumble  dynamics~\cite{berg1972chemotaxis,LOVELY1975477,schnitzer1993theory} (RTD) for a particle moving, for simplicity, in a one dimensional space, with position $x(t)$ is defined by the over-damped Langevin equation
\begin{align}\label{eqn:RT}
\dot{x}(t)=v_0\sigma(t)+\sqrt{2D}\, \xi(t)
\; , 
\end{align}
where $\xi(t)$ is a Gaussian white noise with zero mean and correlations 
$\overline{\xi(t)\xi(t')}=\delta(t-t')$, while $\sigma(t)=\pm 1$ follows a Poisson process, i.e., it is a telegraphic noise with rates $\Gamma(+1\rightarrow-1)=\Gamma_{\downarrow}$ and $\Gamma(-1\rightarrow+1)=\Gamma_{\uparrow}$. The dynamics of $\sigma(t)$ can be solved in terms of the probabilities $P_{\pm}(t)=P(\sigma=\pm 1,t)$, which satisfy
\begin{align}
\dot{P}_+ (t)&=-\Gamma_{\downarrow}P_+(t) +\Gamma_{\uparrow}P_-(t) \; , \\
\dot{P}_-(t) &=\Gamma_{\downarrow}P_+(t) -\Gamma_{\uparrow}P_-(t) \; . 
\end{align}
From these equations we obtain the average $m(t)=\overline{\sigma(t)}=P_+(t)-P_-(t)$, which is
determined by 
\begin{align}
\dot{m}(t) =-(\Gamma_{\uparrow}+\Gamma_{\downarrow}) \, m(t) +(\Gamma_{\uparrow}-\Gamma_{\downarrow})
\; . 
\end{align}
At times, $t\gg 1/\Gamma_+$, with $\Gamma_{\pm}=\Gamma_{\uparrow}\pm\Gamma_{\downarrow}$, we obtain $m_{\infty}=\Gamma_-/\Gamma_+$. The connected correlation $C_{\sigma}(t)$ reads
\begin{align}
C_{\sigma}(t)=\overline{(\sigma(t)-m_{\infty})(\sigma(0)-m_{\infty})}
=
(1-m_{\infty})^2 \, e^{-\Gamma_+t}
\; , 
\end{align}
where we have taken an initial condition $\sigma(0)=1$.
By integrating Eq.~(\ref{eqn:RT}) and using the statistics of the two noises we derive the moments of the particle's position. Focusing on the long-time regime, $t\gg 1/\Gamma_+$, the average position and the particle position variance read as follows:
\begin{align}
\overline{x(t)}&\sim v_0m_{\infty}t \; ,  \\
\overline{\left(x(t)-\overline{x(t)}\right)^2}&\sim 2 D_{\rm RTD} t
\;,
\end{align}
where we have introduced the active diffusion coefficient
\begin{align}\label{eqn:Drtd}
D_{\rm RTD}=D+\frac{v_0^2(1-m^2_{\infty})}{\Gamma_+} = D+\frac{v_0^2 \, [1-(\Gamma_- /\Gamma_+ )^2 ]}{\Gamma_+}
\; . 
\end{align}
We conclude that the diffusion is enhanced by activity, i.e., fluctuations of the velocity due to the telegraph noise.

We then consider the case of Active Brownian Motion~\cite{fily2012athermal,Romanczuk2012active} in one dimension, which is modelled by the stochastic dynamics 
\begin{align}
\dot{x}(t) &=v_0\cos\theta(t)+\sqrt{2D_1} \, \xi(t) \; , \\
\dot{\theta}(t) &=\sqrt{2D_2}\, \eta(t) \; , 
\end{align}
where $\xi(t)$ and $\eta(t)$ are independent Gaussian white noises with zero mean, 
and $\overline{\xi(t)\xi(t')}=\delta(t-t')$ as well as $\overline{\eta(t)\eta(t')}=\delta(t-t')$. To compute the variance of particle's position, we integrate the equations of motion and obtain
\begin{eqnarray}
\overline{\left(x(t)-\overline{x(t)}\right)^2} &=&
v_0^2\int_0^t dt'\int_0^t d\tau \; \overline{\cos(\theta(t'))\cos(\theta(\tau))}
 +
2D_1\int_0^t dt'\int_0^t d\tau \; \overline{\xi(t')\xi(\tau)}
\nonumber\\
&=&
2D_1 t+\frac{v_0^2}{D_2}t+\frac{v_0^2}{D^2_2}\left(e^{-D_2 t}-1\right)
\; , 
\end{eqnarray}
where we have used the fact that $\overline{\cos(\theta(t'))\cos(\theta(\tau))}=e^{-D_2(t-\tau)}\overline{\cos^2(\theta(\tau))}=\frac{1}{2}e^{-D_2(t-\tau)}$. Again,  similar to the case of the 
run-and-tumble particle, we obtain a crossover from short-time diffusion to an intermediate ballistic regime and then to long-time active diffusion with diffusion coefficient given by
\begin{align}
D_{\rm ABP}=D_1+ \frac{v_0^2}{2D_2}
\; . 
\end{align}

Finally, we conclude with the classical Active Ornstein-Uhlenbeck process~\cite{hanggi1994colored,szamel2014self,martin2021statistical}
\begin{align}
\dot{x}(t)&=u(t)+\sqrt{2D} \, \xi(t) \; , \\
\dot{u}(t)&=-\frac{1}{\tau}u(t)+\sqrt{\frac{2D_u}{\tau^2}} \, \eta(t)
\; , 
\end{align}
where  $\xi(t)$ and $\eta(t)$ are independent Brownian motions with zero mean and delta correlations, i.e.,
\begin{align}
     \overline{\xi(t)\xi(t')}=\delta(t-t')
\end{align}
and similarly for $\eta$, i.e., $\overline{\eta(t)\eta(t')}=\delta(t-t')$. In this description, the variable $u(t)$ is the active force or active velocity, which follows an Ornstein–Uhlenbeck process with temporal correlations characterized by a Gaussian colored noise
\begin{align}
\overline{u(t)u(t')}=\frac{D_u}{\tau}\exp(-\vert t-t'\vert/\tau)\,,
\end{align}
with persistence time $\tau$. The integration of the stochastic equations yields the following expression for the variance of particle position
\begin{align}
\overline{\left(x(t)-\overline{x(t)}\right)^2}= 2D t+2D_u\left(t-\tau(1-e^{-t/\tau})\right)\,,
\end{align}
which again displays a crossover from short time diffusive scaling to long-time one, with a renormalized diffusion coefficient
\begin{equation}
D_{\rm AOUP} = D+D_u\,,
\end{equation}
for $t \gg \tau$.

\section{Environment-Assisted Hopping Model }\label{sec:environment_assisted}


In this section we give additional details about the solution of the environment-assisted hopping model discussed in the main text. 
Since the total number of particles $N$ is conserved by the Lindblad evolution, we fix $N=1$ and decompose the density matrix in the single particle sector as
\begin{equation}
\hat\rho = \sum_{i,j \in [1,L]}  \rho_{i,j}|i\rangle \langle j|
\; , 
\label{eq:DecompDensMat1psub}
\end{equation}
with $i$ the lattice site label, $L$ the length of the lattice and 
$|i\rangle$ the eigenstates of the position operator, $\hat x |i\rangle = i |i\rangle$.

For general parameters $ \Gamma_L, \Gamma_R, J $, the equations for the evolution of the coefficients $\rho_{ij}$ for periodic boundary conditions (PBC) read
\begin{eqnarray}
     \dot{\rho}_{ij}(t) 
     &=& 
     - \, {\rm i} J \, \Big( \; \rho_{i + 1,j}(t) + \rho_{i - 1,j}(t) - \rho_{i,j + 1}(t) - \rho_{i,j - 1}(t) \; \Big)  
     \nonumber \\
     && 
     + \, \Gamma_L \, \Big( \, \rho_{i +1,i+1}(t) \, \delta_{ij} - \rho_{ij}(t) \; \Big) 
     \; + \; \Gamma_R \, \Big( \, \rho_{i-1 ,i-1}(t) \, \delta_{ij} - \rho_{ij}(t) \; \Big) 
     \; . 
     \label{eq:generalEqDensMat}
\end{eqnarray}
To obtain the equations ruling the coefficients $\rho_{ij}$ in a problem with open boundary condition (OBC), we simply remove the hopping term between site $L$ and site $1$, both in the Hamiltonian and in the jumps. Equation \eqref{eq:generalEqDensMat} then becomes
\begin{eqnarray}   
        \dot \rho_{ij}(t)&=& 
        - \, {\rm i} J \, \Big( \; \;(1 - \delta_{ij})\rho_{i,j-1}(t) + (1 - \delta_{jL})\rho_{i,j+1}(t) - (1 - \delta_{iL})\rho_{i+ 1,j}(t) - (1 - \delta_{i1})\rho_{i-1,j}(t)\: \: \Big) 
        \nonumber \\
    && + \, \Gamma_L \, \Big( \; \delta_{ij} (1 - \delta_{1j})\rho_{i - 1,i -1}(t) 
    - \frac{1}{2} \Big( (1 - \delta_{1j}) + (1 - \delta_{i1}) \Big) \, \rho_{ij}(t) \; \Big) \; 
    \nonumber \\
    && 
    + \, \Gamma_R \, \Big( \; \delta_{ij}(1 - \delta_{jL})\rho_{i + 1,i +1}(t) -  \frac{1}{2}\Big( (1 - \delta_{jL}) + (1 - \delta_{iL}) \Big) \rho_{ij}(t) \; \Big)
    \; . 
     \label{eq:OBCCoef}
\end{eqnarray}
These equations cannot in general be solved analytically, except in certain limiting cases that we will discuss below. They can, however, be solved numerically and we plot the resulting dynamics in the main text.


We are interested in the dynamics given by equations \eqref{eq:generalEqDensMat}, starting from an initial condition in which the particle is localized at site  $a \in [1, L]$, $    \rho(t = 0) = |a \rangle\langle a|
$. We focus on the variance of the position operator, $\hat{x}=\vert i\rangle\langle i\vert$
\begin{equation*}
    \label{Variance}
    \mbox{Var}(\hat{x}) = \langle \hat{x}^2\rangle - \langle \hat{x}\rangle^2
    \; . 
\end{equation*}
In terms of the density matrix, the first two moments can be expressed as
\begin{align*}
\langle \hat{x}\rangle(t) &= {\rm Tr} (\hat \rho(t) \hat x) = \sum_n \langle n| \sum_{ij} \rho_{i,j}(t) |i \rangle \langle j| \hat x|n \rangle = \sum_i\rho_{ii}(t)i 
\; , \\
    \langle \hat x^2\rangle(t) &= {\rm Tr} (\hat \rho(t) \hat x^2) = \sum_n \langle n | \sum_{ij}\rho_{i,j}(t) |i \rangle \langle j| \hat x^2|n \rangle = \sum_i\rho_{ii}(t)i^2
    \; . 
\end{align*}
The variance is thus given by 
\begin{equation*}
    \mbox{Var}(\hat x) = \sum_i \rho_{i,i}(t) i^2 - \Big(\sum_i \rho_{i,i}(t) i \Big)^2
    \; . 
\end{equation*}

We start showing that the average particle position under PBC acquires simple dynamics. If we take its time derivative and use Eq.~\eqref{eq:generalEqDensMat} we get
\begin{align}
\frac{d\langle \hat{x}\rangle(t)}{dt}=\Gamma_L\sum_i\; i \; \Big(\rho_{i+1i+1}(t)-\rho_{ii}(t)\Big)+
\Gamma_R\sum_i\; i \; \Big(\rho_{i-1i-1}(t)-\rho_{ii}(t)\Big)=\Gamma_-
\end{align}
which describes a drift $\langle \hat{x}\rangle(t) =\Gamma_- \, t$, where we have introduced $\Gamma_{\pm}=\Gamma_L\pm\Gamma_R$.
Similarly, we can write an equation of motion for the second moment which reads
\begin{align}
\frac{d\langle \hat{x}^2\rangle(t)}{dt}=(\Gamma_L+\Gamma_R)+2\Gamma_- \langle \hat{x}\rangle(t)
-\, {\rm i}J\sum_{i} \; (2i+1) \; \Big(\rho_{i,i+1}(t)-\rho_{i+1,i}(t)\Big)
\; . 
\end{align}
To solve this equation we now assume that the off-diagonal matrix elements in Eq.~(\ref{eq:generalEqDensMat}), describing coherences, vanish beyond those close to the diagonal, i.e. $\rho_{i,i+x}=0$ for $x>1$. This allows us to simplify the dynamics of the coherences which now read
\begin{align}
\frac{d\rho_{i,i+1}(t)}{dt}=-{\rm i}J \, \Big(\rho_{i+1,i+1}(t)-\rho_{i,i}(t)\Big)-\Gamma_+\rho_{i,i+1}(t)
\,.
\end{align}
We can now integrate this equation and obtain
\begin{align}
\rho_{i,i+1}(t)= \rho_{i,i+1}(0) e^{-\Gamma_+t}
-{\rm i}J \; e^{-\Gamma_+t}\int_0^t dt' \; e^{\Gamma_+t'} \, \left(\rho_{i+1i+1}(t')-\rho_{ii}(t')\right)
\end{align}
and further drop the first term in the right-hand-side, taking $t\gg 1/\Gamma_+$.
Plugging this result in the equation for the second moment we obtain
\begin{align}
\frac{d\langle \hat{x}^2\rangle(t)}{dt} 
& =
\Gamma_++2\Gamma_- \langle \hat{x}\rangle(t)
-2J^2\int_0^t dt' \; e^{-\Gamma_+(t-t')} \; \sum_{i} \, (2i+1)\left(\rho_{i+1,i+1}(t')-\rho_{i,i}(t')\right)
\nonumber\\
& =
\Gamma_++2\Gamma_-\langle \hat x\rangle(t)+\frac{4J^2}{\Gamma_+}\left(1-e^{-\Gamma_+t}\right)
\label{eqn:dX2}
\end{align}
where in the last step we used the normalization of the density matrix to show that
\begin{align}
\sum_i (2i+1)\Big(\rho_{i+1,i+1}(t)-\rho_{i,i}(t)\Big)=-2
\;. 
\end{align}
After all these steps, solving Eq.~(\ref{eqn:dX2}) we obtain the expression given in the main text for the variance $\mbox{Var}(\hat x)$.

To complete the discussion, we note that the enhancement of diffusivity at long-times is consistent with the Fick's law for particle transport. Indeed, we start from the continuity equation for the particle density, $n_i=c^{\dagger}_ic_i$, which in the single particle sector and for $\Gamma_-=0$ reads
\begin{align}
    \partial_t\langle n_i \rangle =  (J^{c}_{i+1} - J^c_{i})  + (J^{\rm diss}_{i+1} -J^{\rm diss}_i) \label{Supmat:ContinuityEq}
\end{align}
with $J^{c}_i= {\rm i} J\langle c^\dagger_{i + 1}c_i -hc\rangle$ and $J^{\rm diss}_i=\frac{\Gamma_+}{2}\langle n_{i + 1} - n_{i} \rangle$ denoting the local density currents arising from coherent hopping and symmetric incoherent hopping, respectively. Henceforth, we do not write the 
time dependencies explicitly to lighten the notation.
At long times $t\gg 1/\Gamma_+$ one can assume that the off-diagonal matrix elements have already decayed, which leads to the following relation between the two currents
\begin{align}
J^{c}_i=\frac{2J^2}{\Gamma_+}\langle n_{i+1}-n_i\rangle=\frac{4J^2}{\Gamma_+^2}J_i^{\rm diss}
\; .  \label{supmat:CohCurLate}
\end{align}
Plugging in the definition of the total current we find that it reads
\begin{equation}
    J^{\rm tot}_{i + 1}(t) = J^c_{i + 1} + J^{\rm diss}_{i +1} = \frac{1}{2}(\Gamma_+ +  \frac{4 J^2 }{\Gamma_+})\langle  n_{ i + 1} -  n_{i}\rangle = D(J,\Gamma_+) \langle  n_{ i + 1} -  n_{i}\rangle \; . 
\end{equation}
One thus recovers the diffusion coefficient given earlier from the dynamics of particle variance. 

In OBC, an exponential decay of the stationary density, $\rho_{i,i} \propto e^{-i/\xi}$, is expected in systems exhibiting a skin effect. We indeed observe such localization numerically. The stationary distribution follows from~\eqref{Supmat:ContinuityEq} by setting the left-hand-side to zero. Assuming the late-time form of the coherent current given in \eqref{supmat:CohCurLate}, this reduces to: 
\begin{align}
   0 =  \frac{1}{2} \Big(\Gamma_+ + \frac{4 J^2}{\Gamma_+} \Big)\langle n_{i + 1} + n_{i - 1} - 2 n_i\rangle  + \frac{\Gamma_-}{2}  \langle n_{i + 1} - n_{i - 1}\rangle 
   \; . 
   \label{Supmat:ContinuityEqStationnary}
\end{align}
Substituting the Ansatz for the stationary distribution into the equation above yields a second-order polynomial equation for $y = e^{-1/\xi}$: 
\begin{equation}
    0= \Big(D(J, \Gamma_+) + \frac{\Gamma_- }{2}\Big)y^2 - 2D(J, \Gamma_+)y + 
    \Big(D(J, \Gamma_+) - \frac{\Gamma_- }{2}\Big) 
    \; . 
\end{equation}
The discriminant of this equation is simply $\Gamma_-^2$, and the non-trivial solution is $
y = \frac{D(J,\Gamma_+) \pm |\Gamma_-|/2}{D(J,\Gamma_+) + \Gamma_-/2},
$
which leads to
\begin{equation}
    \xi = -1/\log(y) = -\left[\log(\frac{ D(J, \Gamma_+) -  |\Gamma_-|/2}{D(J, \Gamma_+) + \Gamma_-/2 } )\right]^{-1}
    \label{eq:SkinLenght}
\end{equation}
We correctly recover the limits $J \to \infty$ or $\Gamma_- \to 0$, where the localization length diverges, reflecting the suppression of the skin effect, as well as the limit $J \to 0$, where $\xi$ reduces to the Hatano–Nelson localization length, $\xi = -\bigl( 1/\log(\Gamma_L/\Gamma_R)\bigr)$.

\section{Quantum Active Ornstein–Uhlenbeck Process (QAOUP)}

In this Section we provide details on the qAOUP model. 
We start from the Hamiltonian of the system 
\begin{align}
\hat{\mathcal H}_{\rm qAOUP} =\frac{\hat{p}^2}{2m} +\sum_{\alpha}\left(\frac{\hat p_{\alpha}^2}{2m_{\alpha}}+\frac{m_{\alpha}\omega_{\alpha}^2}{2} \hat x_{\alpha}^2\right)+\hat{x}\sum_{\alpha} g_{\alpha}\hat{x}_{\alpha}+\hat{x} \eta(t)
\end{align}
The index $\alpha$ labels the quantum oscillators in the bath.
We take regular Ohmic dissipation~\cite{caldeira1981influence,kamenev2023field} with spectral function $J(\omega)=\pi\sum_{\alpha} \left(g_{\alpha}^2/m_{\alpha}\omega_{\alpha}\right)\, \delta(\omega-\omega_{\alpha})\equiv 4\gamma \omega\exp(-\omega/\omega_c)$. The classical noise 
$\eta(t)$ has zero mean and its time-delayed correlations are $\overline{\eta(t)\eta(t')}\equiv \mathcal{D}(t-t')=\frac{D_a}{\tau}\exp(-\vert t-t'\vert/\tau)$~\cite{martin2021statistical}. We write down the Keldysh action for the quantum particle in terms of classical and quantum components of the particle position, $x_{\rm cl,\rm q}$ and we integrate away the quantum bath degrees of freedom~\cite{kamenev2023field} to obtain
\begin{align}
Z[\eta(t)]=\int D x_{\rm cl}Dx_{\rm q} \; e^{{\rm i} \, S[x_{\rm cl},x_{\rm q};\eta(t)]}\,
\end{align}
with the action $S=S_0+S_{\rm diss}$ written as the sum of two contributions: the first term describes the action of the particle in presence of the classical noise 
\begin{align}
S_0[x_{\rm cl},x_{\rm q}]=
-\int dt \; 
\Big(2mx_{\rm q}(t)\ddot{x}_{\rm cl}(t)+V_{\eta}[x_{cl}(t)+x_q(t)]-V_{\eta}[x_{cl}(t)-x_q(t)]\Big)
\; , 
\end{align}
where $V_{\eta}[x;t]=\eta(t)x$ is the potential associated to the active force, while the second term accounts for the effect of the quantum bath
\begin{align}
S_{\rm diss}[x_{\rm cl},x_{\rm q}]=
\frac{1}{2}\int dt dt' \; \Big(x_q(t)\Sigma^R(t-t')x_c(t')
+x_{cl}(t)\Sigma^A(t-t')x_q(t')\Big)
+\frac{1}{2}\int dt dt'x_q(t)\, \Sigma^K(t-t')\, x_q(t')
\; . 
\end{align}
In this expression, 
$\Sigma^{R/A/K}(t-t')$ are self-energies coming from the Ohmic bath which, in frequency domain, read 
\begin{eqnarray}
\Sigma^{R/A}(\omega)=\int \frac{d\varepsilon}{2\pi}\frac{\varepsilon J(\varepsilon)}{\varepsilon^2-(\omega\pm i0^+)^2} \; , \qquad\qquad\;
\Sigma^K(\omega)= 
\left(\Sigma^R(\omega)-\Sigma^A(\omega)\right)\coth\left(\frac{\beta\omega}{2}\right)
\; . 
\end{eqnarray}
In the expression for $\Sigma^K(\omega)$ we have used the fluctuation-dissipation theorem for the equilibrium bath at temperature $T$. The action above describes a quantum particle in a noisy and dissipative environment. To appreciate the mapping onto the classical AOUP process it is useful to (i) take the low-frequency limit of the bath Green's functions and go back to the time-domain
\begin{align}
\Sigma^R(\omega\rightarrow0)&\sim 2{\rm i}\gamma\omega\rightarrow\Sigma^R(t-t')=2\gamma\delta(t-t')\partial_{t'} \; , \\
\Sigma^K(\omega\rightarrow0)&\sim 8{\rm i} \gamma T \rightarrow\Sigma^K(t-t')
=8 {\rm i} \gamma T\delta(t-t')
\; , 
\end{align}
and (ii) restore the $\hbar$ dependence in the quantum field $x_q/\hbar$ and in the Green's functions. Then, in the semiclassical limit $\hbar\rightarrow0$ one can show that the Keldysh action reduces to
\begin{eqnarray}
S_[x_{\rm cl},x_{\rm q}] 
&=&
-\int dt \; 2x_{\rm q}(t)\Big(m\ddot{x}_{\rm cl}(t)+\gamma \dot{x}_{\rm cl}(t)+\eta(t)\Big)
+4 {\rm i} \gamma T\int dt \, \left(x_q(t)\right)^2
\end{eqnarray}
The quantum vertex $\int dt \, (x_q)^2$ can then be decoupled in terms of an auxiliary stochastic variable with Gaussian correlations, $\overline{\xi(t)\xi(t')}=2\gamma T\delta(t-t')$ which gives 
\begin{eqnarray}
Z[\eta(t)] &=&
\int Dx_cDx_q D\xi \; P(\xi) \; \exp(-2 {\rm i} \int dt \; x_q(t) \Big(m\ddot{x}_{\rm cl}(t)+\gamma \dot{x}_{\rm cl}(t) +\eta(t) -\xi(t) \Big))
\nonumber\\
&=&
\int Dx_c D\xi \; P(\xi) \; \delta\left(m\ddot{x}_c (t) +\gamma \dot{x}_c(t) +\eta(t)-\xi(t)\right)
\end{eqnarray}
where we have ignored irrelevant prefactors, leading to the saddle point equation of motion
\begin{align}
m\ddot{x}_c(t)+\gamma \dot{x}_c(t)+\eta(t)-\xi(t)=0
\end{align}
which in the overdamped regime $m/\gamma\ll 1$ reduces to the AOUP discussed in Sec.~\ref{sec:classical}. At finite $\hbar$ the dynamics of the qAOUP results from the interplay between quantum, thermal and active fluctuations as discussed in the main text and further detailed below.

In the fully quantum regime we can solve the problem using Green's function techniques. We compute the Green's functions of the quantum particle after averaging the Keldysh action over the classical noise $\eta(t)$
\begin{align}
\overline{Z}=
\int D\eta \, P(\eta) \int Dx_{cl}D x_q \; e^{{\rm i} \, S[x_c,x_q;\eta]}=\int D x_{cl}Dx_q\,\exp\left(\frac{1}{2}\sum_{ab=cl,q} \int dt dt' x_a(t)G^{-1}_{ab}(t-t)x_b(t')\right)
\end{align}
where $P(\eta)=e^{-1/2\int dtdt'\; \eta(t)\mathcal{D}^{-1}(t-t')\eta(t')}$ is the probability distribution of the Gaussian AOUP colored noise. The particle position Green's function reads in the frequency domain
\begin{align}
G(\omega)=
\left(
\begin{array}{cc}
    G^K(\omega) &  G^R(\omega) \\
     G^A(\omega)& 0
\end{array}
\right)
\end{align}
from which we can read off directly the retarded/advanced and Keldysh components
\begin{align}\label{eqn:GR}
G^{R/A}(\omega)&=\frac{1}{2m\omega^2+\Sigma^{R/A}(\omega)}\\
\label{eqn:GK}
G^K(\omega)&=-G^R(\omega)G^A(\omega)\left(\Sigma^K(\omega)+4{\rm i} \mathcal{D}(\omega)\right)
\; , 
\end{align}
where we have introduced the Fourier transform of the noise correlator, $\mathcal{D}(\omega)=D_a/\left(1+(\omega\tau)^2\right)$. From this result we confirm that the quantum particle is out of equilibrium, since the AOUP noise $\mathcal{D}(\omega)$ is not associated to any friction. More formally, one can compute the distribution function $F(\omega)$ from the ratio between Keldysh and spectral components and obtain
\begin{align}
F(\omega)\equiv \frac{G^K(\omega)}{G^R(\omega)-G^A(\omega)}=\coth\left(\frac{\beta\omega}{2}\right)+\frac{4\mathcal{D}(\omega)}{J(\omega)}
\end{align}
which is indeed non-thermal. At low-frequency (long-times), however, one can obtain an effective temperature defined as $F(\omega)\sim 2T_{\rm eff}/\omega$ with
\begin{align}
T_{\rm eff}(\omega)=T+\frac{D_a/2\gamma}{1+\omega^2\tau^2}
\end{align}
which is enhanced by the active bath, while at high-frequency (short-times) is given by the equilibrium bath temperature $T$.

Finally, we compute the mean-squared displacement (MSD) of the quantum particle, defined as
\begin{align}
\Delta^2(t)=\overline{\langle (\hat x(t)-\hat x(0))^2\rangle}=
\int \frac{d\omega}{2\pi} \; (1-\cos \omega t) \, {\rm i}G^K(\omega)
\end{align}
where the Keldysh component of the Green's functions, $G^K(t)=-{\rm i}\langle \left\{\hat{x}(t),\hat{x}(0)\right\}\rangle$ is given in Eq.~(\ref{eqn:GK}). We remark that the above MSD is actually a steady-state quantity, not a transient one. To obtain the fully dynamical MSD one would need to solve the Dyson equation for the qAOUP in real-time, which is beyond the scope of this work and does not change qualitatively the picture. Using the Green's functions derived above we obtain
\begin{align}
\Delta^2(t)= \int \frac{d\omega}{2\pi} \; \frac{(1-\cos\omega t)\left(\mbox{Im}\Sigma^K(\omega)+4{\mathcal D}(\omega)\right)}{\left(2m\omega^2\right)^2+4\gamma^2\omega^2}\simeq
\int \frac{d\omega}{2\pi} \; \frac{(1-\cos\omega t)\left(\mbox{Im}\Sigma^K(\omega)+4{\mathcal D}(\omega)\right)}{4\gamma^2\omega^2}
\end{align}
which in the overdamped limit $m/\gamma\ll 1$ reduces to the expression given in the main text. To discuss the different regimes we first consider a finite temperature $T=1/\beta\ll \tau$, for which the Keldysh kernel can be approximated by its zero frequency limit $\mbox{Im}\Sigma^K(\omega\rightarrow0)=8\gamma T$. The integral can then be computed in closed form to obtain
\begin{align}
\Delta^2(t)=2D_T t+2\tilde{D}_a(t +\tau(   e^{- t/\tau} - 1))
\; . 
\end{align}
where $D_T=T/2\gamma$ and $\tilde{D}_a=D_a/4\gamma^2$. This expression coincides with the classical crossover in the variance of the position of an AOUP,  between thermal and active diffusion achieved at times $t\gg\tau$. At very low temperature, or equivalently $\beta\gg 1$, the scaling is different since $\mbox{Im}^K(\omega\rightarrow0)=4\gamma\vert\omega\vert$ and the integral gives 
\begin{align}
\Delta^2(t)=\frac{1}{2\pi\gamma}\mbox{log}(1+\omega^2_ct^2)+2\tilde{D}_a(t +\tau(   e^{- t/\tau} - 1))
\;,
\end{align}
which at intermediate times $t\gg 1/\omega_c$ displays the quantum Brownian motion behavior
\begin{align}
\Delta^2(t) \sim \frac{1}{\gamma}\mbox{log}(\omega_ct)
\;,
\end{align}
while at longer time scales, $t\gg t_*$, the system crosses over to ballistic motion and later to active diffusion for $t\gg \tau$
\begin{align}
\Delta^2(t)= 2\tilde{D}_a t 
\; .
\end{align}
The crossover scale $t_*$ is obtained by comparing the two scaling behaviors, $\frac{1}{\gamma}\mbox{log}(\omega_c t_*)\sim (\tilde{D}_a/2\tau)t_*^2 $, which gives the expression given in the main text using the properties of the Lambert function~\cite{lambert}.

\section{Quantum Active Brownian Motion and Run-and-Tumble dynamics}\label{eqn:spinorbit}

We consider different models of a quantum particle, with position and momentum operators $[\hat x,\hat p]={\rm i}$, coupled to an internal degree of freedom described by a two-level system $\hat \sigma_{a}$, such that $[ \hat\sigma_a, \hat\sigma_b]= {\rm i} \varepsilon_{abc} \hat\sigma_c$. In the first model, the coupling is mediated by the Hamiltonian
\begin{align}\label{eq:s-o}
\hat {\mathcal H}=\frac{ \hat p^2}{2m}+\lambda\hat{p}\hat{\sigma_z}\,,
\end{align}
via spin-orbit coupling with strength $\lambda$. We further assume that the TLS is subject to pump and loss, described by the jump operators 
\begin{align}
\hat L_{\uparrow}=\sqrt{\Gamma_{\uparrow}}\, \hat\sigma_{+}\,, \;\qquad\;\hat L_{\downarrow}=\sqrt{\Gamma_{\downarrow}} \, \hat\sigma_{-}\,,
\end{align}
and that the particle undergoes position diffusion, described the the jump operator $ \hat L_d=\sqrt{\Gamma_d}\hat{p}$ with rate $\Gamma_d$. 
The resulting master equation reads
\begin{align}
\partial_t \hat\rho =- {\rm i} [\hat{\mathcal H}, \hat\rho]+\sum_{\alpha}\hat{L}_{\alpha} \hat\rho \hat{L}^{\dagger}_{\alpha}-\frac{1}{2}\left\{\hat{L}^{\dagger}_{\alpha}\hat{L}_{\alpha}, \hat\rho\right\}
\; . 
\end{align}
From this equation we can derive the equation of motion for the average of a 
generic operator $\langle \hat O\rangle $, which reads
\begin{align}\label{eq:AdjointLindbald}
\frac{d\langle \hat{O}\rangle}{dt}= {\rm i} \langle [\hat{\mathcal H},\hat{O}]\rangle+
\frac{1}{2}\sum_{\alpha}\langle \hat{L}^{\dagger}_{\alpha}[\hat{O},\hat{L}_{\alpha}]\rangle+
\frac{1}{2}\sum_{\alpha}\langle [\hat{L}^{\dagger}_{\alpha},\hat{O}]\hat{L}_{\alpha}\rangle \,.
\end{align}
The system dynamics is therefore described by the following quantities:
\begin{align} \label{eqn:sigma_z}
\frac{d\langle \hat \sigma_z\rangle}{dt}&=\Gamma_- -\Gamma_+\langle \hat \sigma_z\rangle \,, \\
\label{eqn:x}
\frac{d\langle \hat x\rangle}{dt}&=\lambda \langle \hat\sigma_z\rangle+\frac{\langle \hat p\rangle}{m} \; , \\
\label{eqn:x2}
\frac{d\langle \hat x^2\rangle}{dt}&=\frac{1}{m}\langle \hat x \hat p+\hat p \hat x\rangle+2\lambda \langle \hat \sigma_z \hat x\rangle+\Gamma_d \; , \\
\label{eqn:sigmaz_x}
\frac{d\langle \hat\sigma_z \hat x \rangle}{dt}&=\lambda+\frac{1}{m}\langle \hat \sigma_z \hat p\rangle+\Gamma_-\langle \hat x\rangle-\Gamma_+\langle \hat \sigma_z \hat x\rangle 
\; , \\
\label{eqn:x_p}
\frac{d\langle \hat x\hat p \rangle}{dt}&=\frac{\langle \hat p^2\rangle}{m}+\lambda\langle \hat \sigma_z\hat p\rangle\; , \\
\label{eqn:sigmaz_p}
\frac{d\langle \hat\sigma_z \hat p \rangle}{dt}&=\Gamma_-\langle \hat p\rangle-\Gamma_+\langle \hat\sigma_z \hat p\rangle \,,
\end{align}
where we have introduced the total relaxation rate and the net pumping rate $\Gamma_{\pm}=\Gamma_{\uparrow}\pm\Gamma_{\downarrow}$. From the above equations we can obtain the dynamics of the particle position variance, $\mbox{Var}(\hat x)$, which satisfies
\begin{align}
\frac{d\mbox{Var}(\hat x)}{dt}=\Gamma_d+2\lambda\Big(\langle \hat \sigma_z \hat x\rangle-\langle \hat \sigma_z\rangle\langle \hat x\rangle\Big)
+
\frac{2}{m}\Big(\langle \hat x \hat p\rangle-\langle \hat x\rangle\langle \hat p\rangle\Big)=
\Gamma_d+2\lambda\mbox{Cov}(\hat \sigma_z,\hat x)+\frac{2}{m}\mbox{Cov}(\hat x,\hat p)\,,
\end{align}
where $\mbox{Cov}(\hat A,\hat B)=\langle \hat A\hat B\rangle-\langle \hat A\rangle \langle \hat B\rangle$, as given in the main text. The above equations can be integrated in closed form as we are going to discuss. We start by noticing that the momentum $\hat{p}$ is conserved and therefore fix $\langle \hat p\rangle=\langle \hat p\rangle_0$ and $\langle \hat p^2\rangle=\langle \hat p^2\rangle_0$. The dynamics of the TLS is readily obtained from Eq.~(\ref{eqn:sigma_z})
\begin{align}
\langle \hat \sigma_z(t)\rangle =\langle \hat \sigma_z(0)\rangle e^{-\Gamma_+t}+
\frac{\Gamma_-}{\Gamma_+}\left(1-e^{-\Gamma_+t}\right) \,.
\end{align}
We can then solve for the average  particle position 
\begin{align}\label{eqn:x}
\langle \hat x(t)\rangle=\langle \hat x(0)\rangle+\frac{\lambda}{\Gamma_+}\left(\langle \hat\sigma_z(0)\rangle-\frac{\Gamma_-}{\Gamma_+}\right)\left(1-e^{-\Gamma_+t}\right)+v t
\, , 
\end{align}
with the velocity $v=\langle \hat p\rangle_0/m +\lambda \Gamma_-/\Gamma_+$. This result shows that at long times, $t\gg 1/\Gamma_+$, the particle position and velocity are given by the initial ones plus contributions from spin-orbit coupling, while the TLS approaches the steady state on a time scale $1/\Gamma_+$.  We note that for an initial condition in thermal equilibrium, where $\langle \hat p\rangle_0=-m\lambda \langle \hat \sigma_z\rangle_{\rm th}$, the velocity takes the form $v=\lambda\left(\langle \hat \sigma_z\rangle_{\infty}-\langle\hat \sigma_z\rangle_{\rm th}\right)$ and thus vanishes for thermal rates $ \Gamma_-/\Gamma_+$.

We next consider the second moments, which involve correlations between the motional degree of freedom and the TLS. First we solve Eq.~(\ref{eqn:sigmaz_p}) to obtain the dynamics of $\langle \hat{\sigma}_z\hat p\rangle$ which reads
\begin{align}
\langle \hat{\sigma}_z(t)\hat p(t)\rangle=\langle \hat{\sigma}_z(0)\hat p(0)\rangle e^{-\Gamma_+t}+\langle \hat p\rangle_0\frac{\Gamma_-}{\Gamma_+}
\left(1-e^{-\Gamma_+ t}\right)
\label{eq:S70}
\end{align}
and from this we obtain $\langle \{\hat x, \hat p\}\rangle$ solving Eq.~(\ref{eqn:x_p}), 
\begin{align}\label{eqn:xp}
\langle \{ \hat x(t),\hat p(t)\}\rangle=\frac{2\langle \hat p^2\rangle_0}{m}t+\frac{2\lambda}{\Gamma_+}\langle \hat{\sigma}_z(0)\hat p(0) \rangle
\left(1-e^{-\Gamma_+ t}\right)+\langle \hat p\rangle_0\frac{\lambda\Gamma_-}{\Gamma_+}\left( t+\frac{1}{\Gamma_+}(e^{-\Gamma_+t}-1)\right)
\; . 
\end{align}
Next, we consider the dynamics of the correlation between the TLS and the 
position operator, $\langle \hat \sigma_z \hat x\rangle$, which we  obtain
from Eq.~(\ref{eqn:sigmaz_x}) using the results above
\begin{align}\label{eqn:sigmazX}
\langle \hat \sigma_z(t) \hat x(t)\rangle=\langle  \hat \sigma_z(0) \hat x(0)\rangle\,e^{-\Gamma_+t}+
\frac{\lambda}{\Gamma_+}\left(1-e^{-\Gamma_+t}\right)+
\Gamma_-e^{-\Gamma_+t}\int_0^t dt' \; \langle \hat x(t')\rangle \, e^{\Gamma_+t'}+\frac{e^{-\Gamma_+ t}}{m}\int_0^tdt'
\langle \hat{\sigma}_z(t')\hat p(t')\rangle e^{\Gamma_+ t'}\,.
\end{align}
Finally, we can obtain the second moment of particle position,  $\langle \hat x^2\rangle$ whose dynamics from Eq.~(\ref{eqn:x2}) reads
\begin{align}\label{eqn:x2-final}
\langle \hat x^2(t)\rangle=\langle \hat x^2(0)\rangle + \Gamma_d \, t+2\lambda\int_0^{t} dt' \; \langle \hat{\sigma}_z(t')\hat{x}(t')\rangle +\frac{1}{m}\int_0^{t} dt' \; \langle \{ \hat{x}(t'),\hat{p}(t')\}\rangle 
\end{align}
and involves the TLS-particle correlator $\langle \hat \sigma_z \hat x\rangle$. This equation encodes correlations between motion and internal degree of freedom.

\subsection{Symmetric Rates $\Gamma_{\uparrow}=\Gamma_{\downarrow}$}

We discuss first the case of symmetric spin-flip rates, i.e., $\Gamma_-=0$. From Eq.~(\ref{eqn:sigmazX}) and Eq.~(\ref{eq:S70}) we get 
\begin{align}
\langle \hat \sigma_z(t) \hat x(t)\rangle=\langle  \hat \sigma_z(0) \hat x(0)\rangle\,e^{-\Gamma_+t}+
\frac{\lambda}{\Gamma_+}\left(1-e^{-\Gamma_+t}\right)+\frac{\langle  \hat \sigma_z(0) \hat p(0)\rangle}{m}te^{-\Gamma_+t}
\; . 
\end{align}
Plugging this result together with Eq.~(\ref{eqn:xp}) into the expression for the second moment of the particle position we obtain
\begin{align}
\langle \hat x^2 (t) \rangle 
= \; & \; \langle \hat x^2(0)\rangle+\Gamma_d t+\frac{\langle \hat p^2(0)\rangle}{m^2}t^2+\frac{2\lambda^2}{\Gamma_+}
\left(1+\frac{\langle \hat{\sigma}_z(0)\hat{p}(0)\rangle}{\lambda m}\right)\left(t+\frac{1}{\Gamma_+}(e^{-\Gamma_+t}-1)\right)+\nonumber\\
&-\frac{2\lambda}{\Gamma_+}\langle \hat{\sigma}_z(0)\hat{x}(0)\rangle(1-e^{-\Gamma_+t})+
\frac{2\lambda}{m\Gamma_+^2}\langle \hat{\sigma}_z(0)\hat{p}(0)\rangle\Big((1+\Gamma_+ t)e^{-\Gamma_+ t}-1\Big)\,.
\end{align}
If we now consider for simplicity a factorized initial condition between particle and TLS, i.e.
$\langle \hat{\sigma}_z(0)\hat{x}(0)\rangle=\langle \hat{\sigma}_z(0) \rangle \langle \hat{x}(0)\rangle$ and $\langle \hat{\sigma}_z(0)\hat{p}(0)\rangle=
\langle \hat{\sigma}_z(0) \rangle \langle \hat{p}(0)\rangle$,  as well as $\langle \hat x(0)\rangle=\langle \hat p(0)\rangle=0$,  we obtain a compact expression for the particle's position variance, which reads
\begin{align}
\mbox{Var}(\hat x)=\Gamma_d t+\frac{2\lambda^2}{\Gamma_+}\left(t+\frac{1}{\Gamma_+}(e^{-\Gamma_+t}-1)\right)
-\frac{\lambda^2}{\Gamma_+^2}\langle \hat\sigma_z(0)\rangle^2\left(1-e^{-\Gamma_+ t}\right)^2+\frac{\langle \hat p^2(0)\rangle}{m^2}t^2\,.
\end{align}
This expression displays the crossover from diffusion $\bigl[\mbox{Var}(\hat x)\sim \Gamma_dt\bigr]$ to active diffusion $\bigl[\mbox{Var}(\hat x)\sim (\Gamma_d +\frac{2\lambda^2}{\Gamma_+})t\bigr]$ via an intermediate ballistic scaling $\bigl[\mbox{Var}(\hat x)\sim t^2\bigr]$, as reported in the main text, before entering the long-time ballistic regime due to the initial kinetic energy given by the last term. 

\subsection{Asymmetric Rates $\Gamma_{\uparrow}\neq\Gamma_{\downarrow}$}

\begin{figure}[t!]
    \centering
    \includegraphics[width=0.4\linewidth]{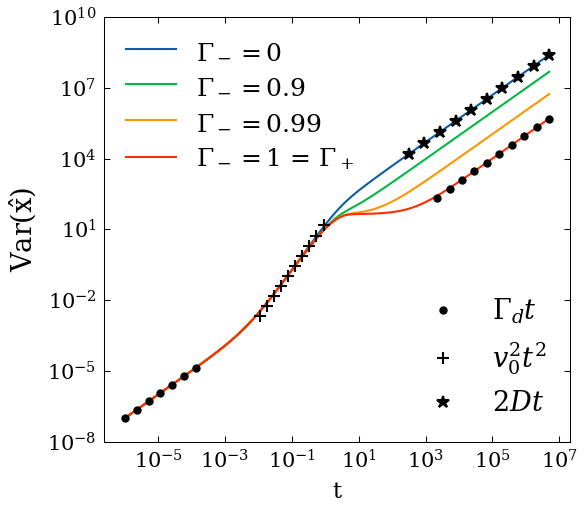}
    \caption{Variance of the particle position for qRTD, in presence of asymmetric rates $\Gamma_-\neq0$. We see the expected crossover from passive diffusion $\bigl(\mbox{Var}(\hat x)=\Gamma_dt\bigr)$ to ballistic dynamics $\bigl(\mbox{Var}(\hat x)=v_0^2t^2\bigr)$ to long-time active diffusion $\bigl(\mbox{Var}(\hat x)=2Dt\bigr)$. While the short-time dynamics is not affected by the asymmetry, the long-time diffusion coefficient decreases with $\Gamma_-$. When $\Gamma_-$ approaches its maximum value ($\Gamma_- = \Gamma_+$), corresponding to fully asymmetric pumping, the enhanced diffusion disappears.   The other parameters are $ \lambda =5, \, \Gamma_+ = 1, \,  \Gamma_d = 0.1$.}
    \label{fig:SupMatFig1}
\end{figure}

We now consider the case of asymmetric rates, $\Gamma_-\neq0$, and we plot in Fig.~\ref{fig:SupMatFig1} the dynamics of the particle variance starting from an uncorrelated initial state and in the infinite mass limit, corresponding to pushing to infinite time the ballistic crossover due to inertia. In particular, we discuss  the evolution for increasing values of $\Gamma_-$. We see that the short-time dynamics, on time scales shorter than $1/\Gamma_+$, is not affected, nor the ballistic crossover at intermediate times. The long-time diffusion coefficients depends instead on $\Gamma_-$, a feature which was absent in the case of the environment-assisted hopping model discussed in Sec.~\ref{sec:environment_assisted}. In particular, we see that increasing the asymmetry the enhancement of diffusivity decreases and eventually in the limit $\Gamma_-=\Gamma_+$, corresponding to a dynamics which only pumps the spin in the up-state, the dynamics recover the passive diffusive scaling. This can be understood intuitively: the diffusion enhancement originates from velocity fluctuations, which vanish when the pumping strongly favors a single spin state ($|\uparrow \, \rangle$ or $|\downarrow \, \rangle$). In this case, the system predominantly occupies a single velocity ($+\lambda$ or $-\lambda$). Interestingly, this feature is also found in the case of the classical RTD discussed in Sec.~\ref{sec:classical}, see Eq.~(\ref{eqn:Drtd}).

\subsection{Stochastic Dynamics from Lindblad unraveling}

In this Section we discuss the unraveling of the Lindblad master equation for the model introduced above. This leads to stochastic quantum trajectories describing the evolution of the system under continuous monitoring~\cite{wiseman2009quantummeasurementand}. We first consider the Quantum Jumps (QJ) unraveling, which takes the form of a stochastic master equation for the unraveled density operator $\rho_{\xi}$:
\begin{align}\label{eqn:qjumps}
d\hat \rho_{\xi} = {\rm i}dt\left(\hat \rho_{\xi} \hat H^{\dagger}_{\rm eff}-\hat H_{\rm eff}\hat \rho_{\xi}\right)+ {\rm i} dt\hat \rho_{\xi}\langle \hat H_{\rm eff}-\hat H^{\dagger}_{\rm eff}\rangle_{\xi}+
\sum_{\alpha}d\xi_{\alpha}\left(\frac{\hat L_{\alpha}\hat \rho_{\xi} \hat L^{\dagger}_{\alpha}}{\langle \hat L^{\dagger}_{\alpha}\hat L_{\alpha}\rangle_{\xi}}-\hat \rho_{\xi}\right)\,.
\end{align}
Here we have defined the effective non-Hermitian Hamiltonian as
\begin{align}
\hat H_{\rm eff}=\hat {\mathcal H}-\frac{{\rm i}}{2}\sum_{\alpha}\hat L^{\dagger}_{\alpha}\hat L_{\alpha}
\;,
\end{align}
and introduced the notation $\langle \hat O\rangle_{\xi}=
\mbox{Tr} \, (\hat \rho_{\xi}\hat O)$ for the average of an operator over the conditional density matrix $\hat \rho_{\xi}$.
Eq.~(\ref{eqn:qjumps}) is stochastic due to the noise increment $d\xi_{\alpha}=\left\{0,1\right\}$ which has Poisson statistics with average
$\overline{d\xi_{\alpha}}=dt\mbox{Tr}\left(\hat \rho \hat L^{\dagger}_{\alpha}\hat L_{\alpha}\right)$, where $\hat \rho = \overline{\hat\rho_{\xi}}$ is the average density matrix which satisfies the Lindblad Master equation~\cite{wiseman2009quantummeasurementand}.
This evolution is therefore composed by a deterministic dynamics driven by $\hat H_{\rm eff}$, for $d\xi_{\alpha}=0$ and by sudden random quantum jumps corresponding to $d\xi_{\alpha}=1$.
From this equation we can derive the stochastic differential equation 
that governs the evolution of a generic  operator $\langle \hat O\rangle_{\xi}$ which reads
\begin{align}\label{eqn:dOqj}
d\langle \hat{O}\rangle_{\xi} = 
{\rm i}dt \left(\langle \hat H_{\rm eff}^{\dagger}\hat O\rangle_{\xi} 
-\langle \hat O \hat H_{\rm eff}\rangle_{\xi} \right)
- {\rm i}dt \langle \hat O\rangle_{\xi} \langle \hat H_{\rm eff}^{\dagger}-\hat H_{\rm eff} \rangle_{\xi} +
\sum_{\alpha}d\xi_{\alpha}\left(\frac{\langle \hat L^{\dagger}_{\alpha} \hat O\hat L_{\alpha}\rangle_{\xi} }{\langle \hat L^{\dagger}_{\alpha}\hat L_{\alpha}\rangle_{\xi}}-\langle \hat O\rangle_{\xi}\right)\,.
\end{align}
Similarly, we consider the quantum-state diffusion (QSD) monitoring protocol~\cite{wiseman2009quantummeasurementand} and the corresponding stochastic master equation
\begin{align}
d\hat \rho_{W}=\left({\rm i} [\hat \rho_{W},\hat {\mathcal{H}}]-\frac{1}{2}\sum_{\alpha}\left\{\hat{L}^{\dagger}_{\alpha}\hat{L}_{\alpha},\hat \rho_{W}\right\}+\hat{L}_{\alpha}\hat \rho_{W}\hat{L}_{\alpha}^{\dagger}\right)dt+
\sum_{\alpha}d W_{\alpha}\left(\hat{L}_{\alpha}\hat \rho_{W}+\hat \rho_{W}\hat{L}_{\alpha}^{\dagger}-\langle \hat{L}_{\alpha}^{\dagger}+\hat{L}_{\alpha} \rangle_W \hat \rho_{W}\right)
\end{align}
where $dW_{\alpha}$ is a real Wiener process such that $\overline{dW}=0$ and $\overline{dW_{\alpha}dW_{\beta}}=dt \, \delta_{\alpha\beta}$, which leads to a stochastic dynamics similar to a Brownian motion in Hilbert space. 
The associated equation of motion for the expectation value of an operator $\langle \hat O\rangle_W=\mbox{Tr} (\hat \rho_{W}\hat O )$ reads 
\begin{align}\label{eqn:dOqsd}
d\langle \hat O\rangle_W&= {\rm i} dt\langle[\hat {\mathcal {H}},\hat O]\rangle_W+
\frac{dt}{2}\sum_{\alpha}\langle \hat{L}^{\dagger}_{\alpha}[\hat O,\hat L_{\alpha}] \rangle_W
+\frac{dt}{2}\sum_{\alpha}\langle [L^{\dagger}_{\alpha},\hat O]\hat L_{\alpha}\rangle_W+\nonumber\\
&+\sum_{\alpha}d W_{\alpha}\left(\langle \hat O\hat L_{\alpha}\rangle_W+\langle \hat L^{\dagger}_{\alpha}\hat O\rangle_W-\langle \hat L^{\dagger}_{\alpha}+\hat L_{\alpha}\rangle_W\langle \hat O\rangle_W\right)
\; . 
\end{align}

Now, we focus on the stochastic dynamics associated to the unraveling of the Lindblad master equation for our model of a particle coupled to a TLS. To keep the notation light in the following we omit the dependence on the noise in the averages of operators in the stochastic dynamics of Eq.~(\ref{eqn:dOqj})-(\ref{eqn:dOqsd}). Using the QSD formulation, the TLS dynamics becomes
\begin{align}
d\langle \hat \sigma_z\rangle= dt\left(\Gamma_--\Gamma_+\langle \hat \sigma_z\rangle\right) +2\Gamma_d dW_d\left(\langle \hat \sigma_z\hat p\rangle-\langle \hat \sigma_z\rangle\langle \hat p\rangle\right)+
\Gamma_{\uparrow}dW_{\uparrow}\langle \hat \sigma_x\rangle\left(1-\langle \hat \sigma_z\rangle\right)-
\Gamma_{\downarrow}dW_{\downarrow}\langle \hat \sigma_x\rangle\left(1+\langle \hat \sigma_z\rangle\right)
\end{align}
while for the particle position operator we obtain
\begin{align}
d\langle \hat x\rangle=dt\left(\langle \hat p\rangle/m+\lambda\langle\hat \sigma_z\rangle\right)+dW_d\left(\langle \hat x\hat p+\hat p\hat x\rangle-2\langle \hat p\rangle\langle \hat x\rangle\right)+\left(\Gamma_{\uparrow}dW_{\uparrow}+\Gamma_{\downarrow}dW_{\downarrow}\right)\left(\langle \hat x\hat \sigma_x\rangle-\langle \hat \sigma_x\rangle\langle \hat x\rangle\right)
\end{align}
where $dW_p,dW_{\uparrow},dW_{\downarrow}$ are the three independent Wiener processes associated to the monitoring of $\hat p$ and $\hat \sigma^{\pm}$. 
On the Bloch sphere we can parametrize $\langle \hat \sigma_z\rangle=\cos\theta$ and write in the overdamped limit
\begin{align}
d\langle \hat x\rangle&=dt\lambda \cos\theta+noise\\
d\theta&=-\frac{dt\left(\Gamma_--\Gamma_+\cos\theta\right)}{\sin\theta}+noise
\end{align}
which is reminiscent of the ABM dynamics of Sec.~\ref{sec:classical} where velocity fluctuations are controlled by an angle $\theta$ performing Brownian motion, with the difference that the quantum noise is multiplicative and there is a deterministic drift term in the $\theta$ dynamics. If the TLS is monitored with quantum jump dynamics, while keeping the QSD monitoring of the particle momentum, we can write instead for the TLS population
\begin{align}
d\langle \hat \sigma_z\rangle= d\xi_+\left(1-\langle \hat \sigma_z\rangle\right)+d\xi_-\left(1+\langle\hat \sigma_z\rangle)\right)
+2\Gamma_d dW_d\left(\langle \hat \sigma_z\hat p\rangle-\langle \hat \sigma_z\rangle\langle \hat p\rangle\right)
\end{align}
while for the particle position
\begin{align}
d\langle \hat x\rangle=dt\left(\langle \hat p\rangle/m+\lambda\langle \hat\sigma_z\rangle\right)+dW_d\left(\langle \hat x\hat p+\hat p\hat x\rangle-2\langle \hat p\rangle\langle \hat x\rangle\right)
\end{align}
which in the overdamped limit is again reminiscent of the RTD of Sec.~\ref{sec:classical}, since now the TLS dynamics performs random switching between $\langle \hat \sigma_z\rangle=\pm1$ corresponding to a telegraph process, with the additional multiplicative monitoring of momentum diffusion.

\begin{figure}[t!]
    \centering
    \includegraphics[width=0.4\linewidth]{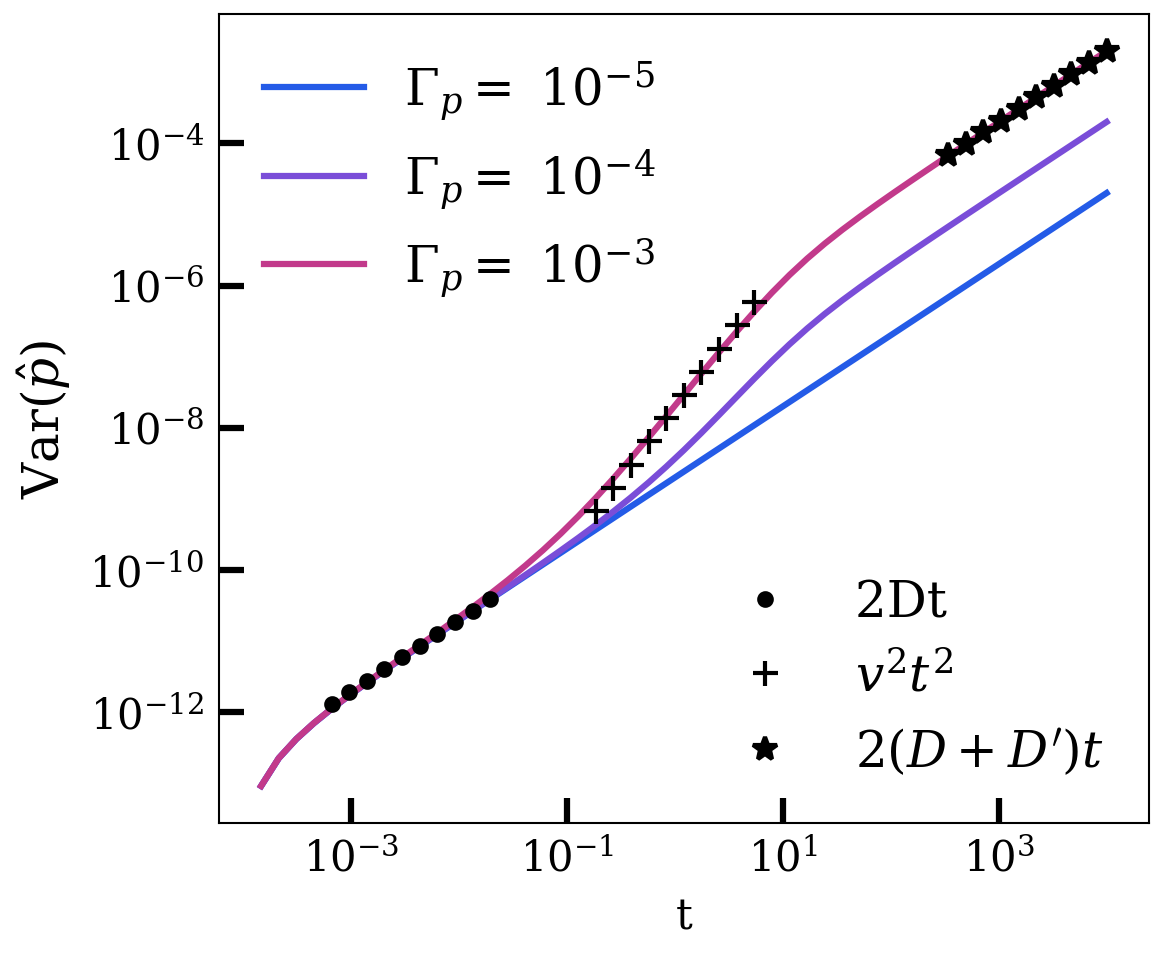}
    \caption{Variance of the particle momentum for the dissipative spin-orbit coupling model discussed in Sec.~\ref{eqn:diss_so}.
    We consider the symmetric limit $\Gamma_L = \Gamma_R = \Gamma $ for different values of $\Gamma_p$. Parameters :  $[\Gamma, p_0, m] = [0.1, 10^{-2},1]$ . 
    }
    \label{fig:SupMatFig2}
\end{figure}
\section{Active Dynamics from Dissipative Spin-Orbit Coupling}\label{eqn:diss_so}

In this Section we introduce a fully dissipative version of the model described in Sec.~\ref{eqn:spinorbit}, where the coupling between motional degree of freedom and TLS is implemented via jump operators. Specifically, we consider a free particle
\begin{align}
\hat{\mathcal{H}}=\frac{\hat p^2}{2m}
\end{align}
subject to a dissipative evolution that couples internal and motional degrees of freedom via the jump operators
\begin{align}\label{eqn:jump_so}
\hat L_L=\sqrt{\Gamma_L}\hat \sigma_-e^{ip_0 \hat x}\,,
\qquad
\hat L_R=\sqrt{\Gamma_R}\hat \sigma_-e^{-ip_0 \hat x} \,,
\end{align}
together with spin pumping and momentum diffusion
\begin{align}
\hat L_s=\sqrt{\Gamma_{\uparrow}}\hat \sigma_+\,\qquad\,
\hat L_d=\sqrt{\Gamma_p}\hat x \,.
\end{align}
The jump operators in Eq.~(\ref{eqn:jump_so}) implement an effective spin-orbit coupling, since every time that the TLS is flipped down, the particle momentum acquires a shift $\pm p_{0}$ implemented by the operator $e^{\pm ip_0 \hat x}$. This interaction arises naturally, in a coherent version, when discussing the effect of recoil of a classical laser beam on the atom position. To implement it dissipatively one could for example embed the driven atom in a leaky cavity and eliminate the cavity mode, obtaining an effective dissipation for the atom.
By using Eq.~\eqref{eq:AdjointLindbald} we can derive the equations of motion for this system under Lindblad dynamics.
We first compute the dynamics of the TLS which reads 
\begin{align}
\frac{d\langle \hat \sigma_z\rangle}{dt}=\left(\Gamma_{\uparrow}-\Gamma_L-\Gamma_R\right)-\left(\Gamma_{\uparrow}+\Gamma_L+\Gamma_R\right)\langle \hat \sigma_z\rangle\,,
\end{align}
where the total relaxation rate is now $\Gamma_{\uparrow}+\Gamma_L+\Gamma_R$. The evolution of the average momentum is given by
\begin{align}
\frac{d\langle \hat p\rangle}{dt}=p_0\left(\Gamma_L -\Gamma_R\right)\langle \hat \sigma_+\hat \sigma_-\rangle=\frac{p_0}{2}\left(\Gamma_L -\Gamma_R\right)
\left(1+\langle \hat \sigma_z \rangle\right)\,.
\end{align}
We see that due to the momentum kicks $p_0$ imposed by the laser, the particle momentum has a drift term controlled by the excited state population and by the asymmetry $\Gamma_L-\Gamma_R$. Similarly, the dynamics of the momentum fluctuations read
\begin{align}
\frac{d\langle \hat p^2\rangle}{dt}=\frac{1}{2}p_0^2\left(\Gamma_L+\Gamma_R\right)\left(1+\langle \hat \sigma_z\rangle\right)+
p_0\left(\Gamma_L -\Gamma_R\right)\left(\langle \hat p\rangle +\langle \hat \sigma_z\hat  p\rangle\right)+\Gamma_p\,.
\end{align}
To obtain this result we have used the following identities
\begin{align}
[\hat \sigma_+e^{-ip_0\hat x},\hat p^2]&=\hat \sigma_+\left(p_0^2+2\hat pp_0\right)e^{-ip_0\hat x} 
\; , \\
[\hat p^2,\hat \sigma_-e^{ip_0\hat x}]&=\hat \sigma_- e^{ip_0\hat x}\left(p_0^2+2p_0\hat p\right)\; .
\end{align}
We see that in the case of asymmetric couplings, the dynamics of $\langle \hat p^2\rangle$ is coupled to a mixed correlator between internal
and motional degree of freedom, $\langle \hat \sigma_z\hat p\rangle$. We can therefore evaluate its dynamics to get
\begin{align}
\frac{d\langle \hat \sigma_z\hat p\rangle}{dt}
=
-\frac{\left(\Gamma_L-\Gamma_R\right)}{2}p_0\left(1+\langle \hat \sigma_z\rangle\right)+
\Gamma_{\uparrow}\left(\langle\hat  p\rangle-\langle \hat \sigma_z\hat p\rangle\right)
\; . 
\end{align}
We can therefore integrate this dynamics and obtain the evolution of the momentum fluctuations to fully solve the problem. In the symmetric case, $\Gamma_L=\Gamma_R=\Gamma$ the variance of $\hat p$ is given by
\begin{align}
     \mbox{Var}(\hat p ) =\mbox{Var}(\hat p)_0+2 \Gamma p_0^2\left(1+\langle \hat \sigma_z \rangle _\infty\right)t +
     2 \Gamma p_0^2\,\frac{(\langle \hat \sigma_z \rangle _{\infty} - \langle \hat \sigma_z \rangle _0)}{\Gamma_{\rm tot}}e^{- \Gamma_{\rm tot} t}  
\end{align}
with $\Gamma_{\rm tot} = 2\Gamma + \Gamma_d.$ We recover diffusive behavior at short times, governed by the initial density of spin-up particles
\begin{align}
    \mbox{Var}(\hat p) \sim2D t + v^2 t^2 + \mathcal{O}(t^3) \,,
\end{align}
where we have defined
\begin{align}
D = \Gamma p_0^2 \left(1+\langle \hat \sigma_z\rangle_0\right)\,, \; \quad v^2 = \Gamma \Gamma_{\rm tot} p_0^2 (\langle \sigma_z\rangle _0  - \langle \sigma_z\rangle _\infty )\,.
\end{align}
The ballistic contribution arises from the mismatch between the initial spin distribution and the stationary distribution. At long times, the diffusion coefficient is instead determined by the stationary density of spin-up particle.
\begin{align}
    \mbox{Var}(\hat p) \sim 2(D + D') \;t
\end{align}
where the enhancement in the diffusion coefficient reads  $D' = \Gamma p_0^2( \langle \sigma_z \rangle_{\infty} -\langle \sigma_z\rangle_{0}) /2 $. The resulting phenomenology is again similar to classical models of active particles, where now, as opposed to previous cases, diffusion happens in momentum space rather than in real space. We plot the crossover in the momentum variance in Fig.~\ref{fig:SupMatFig2}.


%